# Cu-In Halide Perovskite solar absorbers

Xin-Gang Zhao[1], Dongwen Yang[1], Yuanhui Sun[1], Tianshu Li[1], Lijun Zhang[1],*, Liping Yu[2], and Alex Zunger[3]

[1]State Key Laboratory of Superhard Materials, Key Laboratory of Automobile Materials of MOE, and College of Materials Science and Engineering, Jilin University, Changchun 130012, China

[2]Department of Physics, Temple University, Philadelphia, PA 19122, USA

[3]University of Colorado, Renewable and Sustainable Energy Institute, Boulder, Colorado 80309 USA

**Abstract:** The long-term chemical instability and the presence of toxic Pb in otherwise stellar solar absorber $APbX_3$ made of organic molecules on the $A$ site and halogens for $X$ have hindered their large-scale commercialization. Previously explored ways to achieve Pb-free halide perovskites involved replacing $Pb^{2+}$ with other similar $M^{2+}$ cations in $ns^2$ electron configuration, *e.g.*, $Sn^{2+}$ or by $Bi^{3+}$ (plus $Ag^+$), but unfortunately this showed either poor stability ($M$ = Sn) or weakly absorbing oversized indirect gaps ($M$ = Bi), prompting concerns that perhaps stability and good optoelectronic properties might be contraindicated. Herein, we exploit the electronic structure underpinning of classic Cu[In,Ga]Se$_2$ (CIGS) chalcopyrite solar absorbers to design Pb-free halide perovskites by transmuting 2Pb to the pair $[B^{IB} + C^{III}]$ such as [Cu + Ga] or [Ag + In] and combinations thereof. The resulting group of double perovskites with formula $A_2BCX_6$ ($A$ = K, Rb, Cs; $B$ = Cu, Ag; $C$ = Ga, In; $X$ = Cl, Br, I) benefits from the ionic, yet narrow-gap character of halide perovskites, and at the same time borrows the advantage of the strong Cu($d$)/Se($p$)→Ga/In($s/p$) valence-to-conduction-band absorption spectra known from CIGS. This constitutes a new group of *CuIn–based Halide Perovskite* (CIHP). Our first-principles calculations guided by such design principles indicate that the CIHPs class has members with clear thermodynamic stability, showing direct band gap, and manifesting a wide-range of tunable gap values (from zero to about 2.5 eV) and combination of light electron and heavy-light hole effective masses. Materials screening of candidate CIHPs then identifies the best-of-class Rb$_2$[CuIn]Cl$_6$, Rb$_2$[AgIn]Br$_6$, and Cs$_2$[AgIn]Br$_6$, having direct band gaps of 1.36, 1.46, and 1.50 eV, and theoretical spectroscopic limited maximal efficiency comparable to chalcopyrites and CH$_3$NH$_3$PbI$_3$. Our finding offers new routine for designing new-type Pb-free halide perovskite solar absorbers.

1. Introduction

We identify theoretically a new promising group of Pb-free halide perovskites that has direct band gaps spanning the solar range, is thermodynamically resilient to decomposition, has low electron and combined light-heavy hole effective masses and a rather strong light absorption near threshold, and could thus replace Pb-based hybrid materials as solar absorbers. This group of materials is designed by combining the theoretical understanding of (i) the factors that limited the performance of some of the previously considered halide perovskites (where Pb was replaced by other related elements)[1–9] with (ii) the special features that enabled the high performance of Cu-based ternary chalcopyrites (Cu(In,Ga)Se$_2$, CIGS) as photovoltaic absorbers.[10–12] The significance of this work is stepping out of conventional design principles of replacing $Pb^{2+}$ with other similar n$s^2$ cations and considering instead transmuting two $Pb^{2+}$ to the pair of a group IB (Cu$^+$ and Ag$^+$) and a group III (Ga$^{3+}$ and In$^{3+}$) cation, exemplified by Cs$_2$[AgIn]Cl$_6$. Not only are these halide perovskites free of unwanted toxic Pb, or easily oxidized Sn replacement of Pb,[1,2] as well as avoiding the Ag + Bi transmutation that causes indirect and oversized band gaps,[5,6,13] but they also benefit from the $d^{10}$ electronic motif dominated valence bands underlying the successful Cu-based chalcopyrites that enable a rather strong absorption curve and promise favorable doping and good materials stability.[10,14,15]

***Rapid progress in $APbX_3$ group hybrid perovskite solar cells research and the challenges it raises.*** The success of the long-ago discovered[16] but until recently unappreciated hybrid halide perovskites of the $APbX_3$ group (left panel of Fig 1a) as superior solar absorbers rapidly reaching a power conversion efficiency of 22%[17–25] from initial value of 3.8%[17] has focused greater attention of the photovoltaic community on the need for understanding-based deliberate design and discovery of novel solar absorbers. Our developing understanding of the key properties behind the success of $APbX_3$ include its (i) very strong and fast-rising direct-gap optical transition between valence $Pb(s)/X(p)$ and conduction $Pb(p)$ states,[26] (ii) low exciton binding energy[27,28] allowing fast disengagement of optically generated electrons from holes, (iii) simultaneously light effective masses of electron and hole facilitating their transport (*e.g.*, ultralong diffusion length),[29] (iv) energetically shallow intrinsic defect levels beneficial to bipolar conductivity and meanwhile minimizing carrier trapping and scattering,[30–32] and (v) last but not least, suitability of low-cost, non-vacuum solution-preparation routes for growing films. Despite enormous success of this class of materials, major challenges have been posed by (a) the toxicity of Pb and (b) the general instability of $APbX_3$ under various conditions,[33–38] *e.g.*, rising from thermal loss of halogen (*e.g.*, as HX) at relatively low temperatures,[39,40] and from the decomposition reaction $APbX_3$→$AX$ + $PbX_2$ that is slightly exothermic for A = CH$_3$NH$_3^+$.[3,40]

***Previous proposed single-substitution solutions to the challenge of $APbX_3$.*** Solutions to these challenges in $APbX_3$ were naturally first sought by substitution of the sites A, Pb or X[41,3,42] or even alloying of various isovalent species on the same site.[43–45,19] Experimentally, increasing the band gap (as needed for tandem cells) by replacing I with Br lead to a curious light-induced instability,[19] whereas replacing organic cations by Cs limited the cell efficiency to ~10%.[46,47] Whereas the decomposition tendency could be slowed down by replacing A = CH$_3$NH$_3^+$ with the larger molecules such as CH$_2$(NH$_3$)$_2^{+43}$ or some alkali cations (Cs$^+$, Rb$^+$)[48,49] having now an endothermic reaction enthalpy, this offered but a partial solution to the instability problem. Replacing Pb by Sn in

APbX$_3$ gives a maximum solar cell efficiency of only 6%, accompanying with remarkably low (~0.25 ns) carrier lifetime.[1,2] The low conversion efficiency of ASnX$_3$ might be attributed to a consequence of the defect physics on the multi-valency nature of Sn (stable as both 4+ and 2+), *i.e.*, leading to the formation of deep defect levels as carrier-trapping centers. Indeed, this is analogous to the detrimental role of multi-valent Sn in Cu$_2$[ZnSn]S$_4$ (CZTS) used as a replacement for CuInSe$_2$.[50] Theoretical screening of the compounds where Pb in APbX$_3$ were replaced by other isovalent elements reveals that in most cases this results in non-ideal band gaps.[42,51]

***Strategy of double substitution of Pb sites in double-perovskite structure utilizing Bi.*** An alternative approach previously attempted to design Pb-free halide perovskites has been to replace two Pb$^{2+}$ ions in single perovskite APbX$_3$ with an ion pair of a monovalent B$^+$ and a trivalent C$^{3+}$ in double perovskite A$_2$[BC]X$_6$ (Fig. 1a). For example, the 2Pb$^{2+}$ was transmuted into [Ag$^+$ + Bi$^{3+}$], generating Bi-based double-perovskites Cs$_2$[AgBi]Cl$_6$ or Cs$_2$[AgBi]Br$_6$ (Fig. 1b).[6,5,13] Unfortunately, in contrast with APbX$_3$ having a direct band gap at the R point of the Brillouin zone,[32] Cs$_2$[AgBi]Cl$_6$ has an indirect gap between the valence band maximum (VBM) at X and the conduction band minimum (CBM) at L (nearly degenerate with the Γ state) (Fig. 1b).[7,13] Instead of the (anti-bonding) coupling between the upper Pb(6$s$) orbital and the deeper X($p$) orbital in the valence band of APbX$_3$, we have in the valence band of Cs$_2$[AgBi]Cl$_6$ a coupling between the *orbitally asymmetric cationic framework* Ag($d$) + Bi(6$s$) that interacts with the anionic Cl($p$) states, placing the VBM off-center at the X point, and making the gap indirect.[7,52] The ensuing gap values in Cs$_2$[AgBi]Cl$_6$ or Cs$_2$[AgBi]Br$_6$ are above 2.0 eV,[5,6] unsuitable as solar absorbers. Use of a cationic complex [B + C] made of *s*-orbital components alone such as [Tl$^+$ + Bi$^{3+}$] indeed produces direct band gap as in [Pb$^{2+}$ + Pb$^{2+}$] as predicted[4,7] and verified experimentally[4] in (CH$_3$NH$_3$)$_2$[TlBi]Br$_3$. Unfortunately, such systems contain another toxic element (Tl) and has a gap of 2.16 eV[4] that is too high for single-junction solar cells.

***Lessons distilled from previous understanding of CuIn-based chalcopyrite solar absorbers.*** The I-III-VI$_2$ chalcopyrites Cu[In,Ga]Se$_2$ is a classical solar absorber generated by exploiting the idea of cations transmutation, *i.e.*, via conversion of 2Zn$^{2+}$ in zinc blende ZnSe to cation pair of [Cu$^+$ + (In/Ga)$^{3+}$] (Fig. 1c). The band structure illustrated for a related chalcopyrite member (AgInSe$_2$ in Fig. 1c) is distinguished by having a closed (Cu/Ag) $d^{10}$ shell which dominates the valence bands, light absorption, defect properties, and carrier dynamics.[53,54] Optical absorption of CIGS due to Cu($d$)/Se($p$)→Ga/In($s$/$p$) valence-to-conduction-band transition is direct and strong, reaching its maximal value in a narrow energy window above threshold, thus the film can be thin (unlike Si) and drift diffusion is enabled. Because of the anti-bonding hybridization between Cu($d$) and Se($p$) orbitals dominating the valence bands, Cu vacancies as acceptors producing holes are energetically shallow,[54,55] with concentration that is controllable via growth. Specifically, the existence of Cu-poor regions (thus hole-rich) and Cu stoichiometric regions (electron rich) naturally creates spatially separate channels of transport for electrons and holes,[56–58] leading to weak carrier recombination and good diffusion length. These advantages made CIGS achieve power conversion efficiency of 22% comparable to MAPbI$_3$.[25] But its growth and processing methods (often vacuum related) are not as accessible as the low temperature solution growth used for halide perovskites.

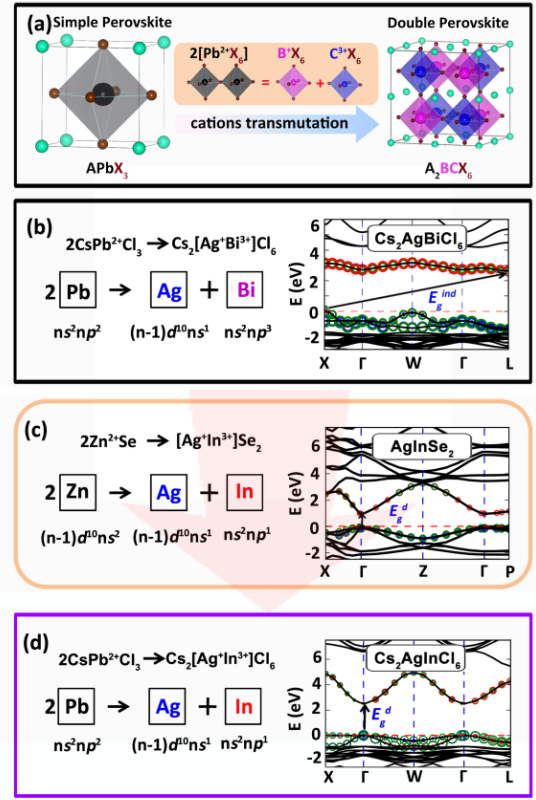

**Figure 1.** (a) Illustration of cations transmutation strategy (by converting 2Pb$^{2+}$ to pair of [B$^+$ + C$^{3+}$]) to design Pb-free halide double perovskites. (a, b, c) Materialization of Bi-based double-perovskite (Cs$_2$[AgBi]Cl$_6$), chalcopyrite (AgInSe$_2$), and CuIn–based Halide Perovskite (CIHP) (Cs$_2$[AgIn]Cl$_6$) via cations transmutation, and their electronic band structures. Circles with different sizes represent ortibal projection of band edge states (with red for In($s$/$p$) or Bi($s$/$p$), blue for Ag($d$), and green for Cl($p$)). The band gap feature is indicated by $E_g^d$ for direct and $E_g^{ind}$ for indirect.

***Design idea of CuIn–based Halide Perovskites.*** The discussion above underlies the concept that to assure a direct gap of double perovskite A$_2$[BC]X$_6$, the cation pair [B + C] better has either the same cation orbital symmetry, or have one high-lying cation orbital unobstructed by repulsion from the deeper cation orbital of the other site. Inspired by this understanding as well as the cations transmutation scheme of classical solar chalcopyrites (i.e., 2Zn$^{2+}$ → Cu$^+$ + Ga$^{3+}$/In$^{3+}$), we herein propose to design Pb-free halide double perovskites of A$_2$[BC]X$_6$ via transmuting 2Pb$^{2+}$ to cation pair of [B + C] with B = (Cu$^+$/Ag$^+$) and C = (Ga$^{3+}$/In$^{3+}$) (Fig. 1d). We build upon the understanding gained from the mechanisms of chalcopyrites being high-efficiency absorbers, and offer through first-principle calculations a new group of CuIn–based Halide Perovskite (CIHP). We consider elemental constitutions of A$^I$ = K, Rb, Cs; B$^{IB}$ = Cu, Ag; C$^{III}$ = Ga, In; X$^{VII}$ = Cl, Br, I for the A$_2$[BC]X$_6$ CIHPs, totally 36 candidate compounds. The significant features of band structure of CIHPs (represented by Cs$_2$[AgIn]Cl$_6$ in Fig. 1d) show desired direct-gap nature, overcoming the indirect-gap problem of Cs$_2$[AgBi]Cl$_6$ and Cs$_2$[AgBi]Br$_6$. Resembling chalcopyrites, the CIHPs have the Cu/Ag($d$)-X($p$) hybridization dominated valence bands offering a rather strong Cu($d$)/X($p$)→Ga/In($s/p$) valence-to-conduction-band optical transition. This class of Pb-free CIHPs has a wide-range tunable direct band gaps ranging from zero to about 2.5 eV, as well as low electron and combined light-heavy hole effective masses. We identify via materials screening process six CIHPs showing simultaneous thermodynamic and dynamic phonon stability. Among them Rb$_2$[CuIn]Cl$_6$, Rb$_2$[AgIn]Br$_6$, and Cs$_2$[AgIn]Br$_6$, show direct band gaps of 1.36, 1.46, and 1.50 eV, and theoretical solar cell efficiency comparable to chalcopyrites and CH$_3$NH$_3$PbI$_3$.

## 2. Computational Methods

Our first-principles calculations are carried out by using plane-wave pseudopotential approach within density functional theory (DFT) as implemented in the Vienna Ab Initio Simulation Package (VASP).[59,60] The electron-core interactions are described with the projected augmented wave pseudopotentials[60] with (n-1)$s^2$(n-1)$p^6$n$s^1$ for K/Rb/Cs, (n-1)$d^{10}$n$s^1$ for Cu/Ag, n$s^2$n$p^1$ for Ga/In, and n$s^2$n$p^5$ for Cl/Br/I as valence electrons. The generalized gradient approximation formulated by Perdew, Burke, and Ernzerhof (PBE)[61] is used as exchange correlation functional. We adopt the standard cubic double-perovskite or Elpasolite structure (in space group of $Fm\text{-}3m$) for all A$_2$[BC]X$_6$. Our explorative calculations indicate this structure has the lowest energy among all the structures in A$_2$[BC]X$_6$ stoichiometry with different arrangement patterns of BX$_6$ and CX$_6$ octahedra (Supplementary Fig. S1). This is in accord with the experimentally established structure of Bi-based A$_2$[BC]X$_6$ perovskites.[5,6] Structures are locally optimized (by relaxing both lattice parameters and internal atomic coordinates of the preassigned $Fm\text{-}3m$ space group) via total energy minimization. The optimized kinetic energy cutoffs deciding the size of the plane-wave basis set and the k-points mesh with grid spacing of less than 2π×0.10 Å$^{-1}$ are used to ensure the residual forces on atoms converged to below 0.0002 eV/Å. To reduce the self-interaction error of DFT in band gaps calculations, we used the Heyd-Scuseria-Ernzerhof (HSE) hybrid functional approach with standard 25% exact Fock exchange included.[62] For the winning CIHP compounds, the HSE functional is used both for structural optimization and for evaluating the band gap at the optimized geometry. This largely solves the band gap underestimation problem underlying DFT. Benchmark calculations on known chalcopyrites and Bi-based A$_2$[BC]X$_6$ (Supplementary Fig. S2) indicate that our approach gives rather small difference between theory and experiment for lattice constants, and provides correct trends of band gap values (with a small underestimation of 0.3 eV) for chalcopyrite series and shows good experiment-theory agreement of gap values for Bi-based A$_2$[BC]X$_6$. The effect of spin-orbit coupling (SOC) on the electronic structure of the representative compound Cs$_2$[AgIn]Br$_6$ has been tested and found to be negligible on band-edge electronic structure and thus band gap and carrier effective masses. This is shown in Supplementary Fig. S3. We conclude that the SOC can be reasonably neglected in considering the leading features of optoelectronic properties of the CIHP compounds. Harmonic phonon spectrum is calculated with a finite-difference supercell approach implemented in Phonopy code,[63] and room-temperature phonon spectrum is obtained by taking into account anharmonic phonon-phonon interaction with a self-consistent ab initio lattice dynamical method.[64] Carrier effective masses are calculated via second derivative of band dispersion E($k$) from the HSE band structure calculations. To evaluate material-intrinsic solar cell efficiency of winning CIHPs, the "spectroscopic limited maximum efficiency (SLME)" based on the improved Shockley-Queisser model[65] is calculated. Creation of calculations, extraction of calculated results, and post-processing analysis are performed by using an open-source Pyhon infrastructure designed for large-sale high-throughput energetic and property calculations of functional materials, in in-house developed Jilin University Materials-design Python Package (Jump$^2$, to be release soon). More detailed computational procedures are described in Supplementary Sec. I.

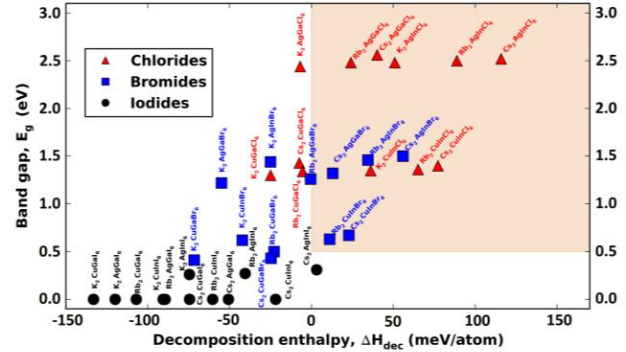

**Figure 2.** Calculated band gap ($E_g$) plotted vs decomposition enthalpy ($\Delta H_{\text{dec}}$, see text) for all the candidate CIHPs. The criterion for materials screening of stable solar compounds, i.e., $\Delta H_{\text{dec}} > 0$ and $0.5 < E_g < 3.0$, is shaded.

## 3. Results and Discussion

*Materials screening based on stability against decomposition and solar band gap values*: The brief history of solar absorbing perovskites has indicated that stability matters.[39,40,66] Using the paradigm of CIHPs based on electronic structure consideration must hence be supplemented by analysis of thermodynamic stability. We use two stability criteria for materials screening, the *first* being simpler and hence readily applicable to a larger group of materials (together with appropriate band gaps used as an initial screening filter) and the *second* being more rigorous and computationally costly and thus applied to the compounds passing the initial filter. The initial stability filter is chemical stability against decomposition reaction into common binary compounds (A$_2$BCX$_6$ → 2AX + BX + CX$_3$) characterized by the decomposition enthalpy ($\Delta H_{\text{dec}}$, defined via free energy difference after and before the above reaction, $\Delta H = 2E_{AX} + E_{BX} + E_{CX_3} - E_{A_2BCX_6}$). The positive $\Delta H_{\text{dec}}$ means reaction being endothermic resulting in suppressed decomposition of A$_2$BCX$_6$. The more rigorous stability filter, involves examination of all other decomposition channels into various combinations of competing phases.[67,68]

Fig. 2 shows the calculated primary stability metric $\Delta H_{\text{dec}}$ plotted vs calculated band gaps $E_g$ (all being direct) for the candidate CIHPs (the explicit data are listed in Supplementary Table S1). The $E_g$ corresponds to the single-phase compound and might be somewhat underestimated value as can be judged from tests on known compounds in Supplementary Fig. S2. One observes a general trend that compounds with the larger $E_g$ have the higher stability with respect to the above simple decomposition reaction. Band gap values span a broad range from zero in metallic iodides to over 2.5 eV of several chlorides. Generally, the Ag-based CIHPs show the larger gaps than those of the Cu-based. Target gaps may be needed for high-efficiency tandem solar cells, spanning three ranges: 1.2-1.3 eV, 1.7 eV, and 2.2-2.4 eV. Adoption of the screening filter of $\Delta H_{dec} > 0$ and $0.5 < E_g < 3.0$ (shaded region in Fig. 2) lead us to select 13 CIHPs as tentative stable solar materials. These include eight chlorides (*i.e.*, three Cu-based ones of $K_2[CuIn]Cl_6$, $Rb_2[CuIn]Cl_6$, and $Cs_2[CuIn]Cl_6$ and five Ag-based ones of $Rb_2[AgGa]Cl_6$, $Cs_2[AgGa]Cl_6$, $K_2[AgIn]Cl_6$, $Rb_2[AgIn]Cl_6$, $Cs_2[AgIn]Cl_6$) and five bromides (*i.e.*, $Rb_2[CuIn]Br_6$, $Cs_2[CuIn]Br_6$, $Rb_2[AgIn]Br_6$, $Cs_2[AgGa]Br_6$, and $Cs_2[AgIn]Br_6$). It should be noted that although we use the selection criterion of $\Delta H_{dec} > 0$, we believe that somewhat metastable structures with negative $\Delta H_{dec}$ might still be formable as the halide bonds are strong enough to withstand moderate metastability.

Supplementary Sec. II discusses formability of perovskite structure of candidate CIHPs from the classical point of view of close packing via the Goldschmidt tolerance factor $t$ and the octahedral factor $\mu$ using the idealized solid-sphere model. We find that of the stable CIHPs satisfying the above noted first principles DFT primary stability metric $\Delta H_{dec} > 0$, only 64% meet the classic formability criterion, so we abandon the latter argument as being insufficiently accurate. A possible factor responsible for this inconsistency is the reliability of evaluation of $t$ and $\mu$ in current double-perovskite system containing mixed cations at the octahedral site.

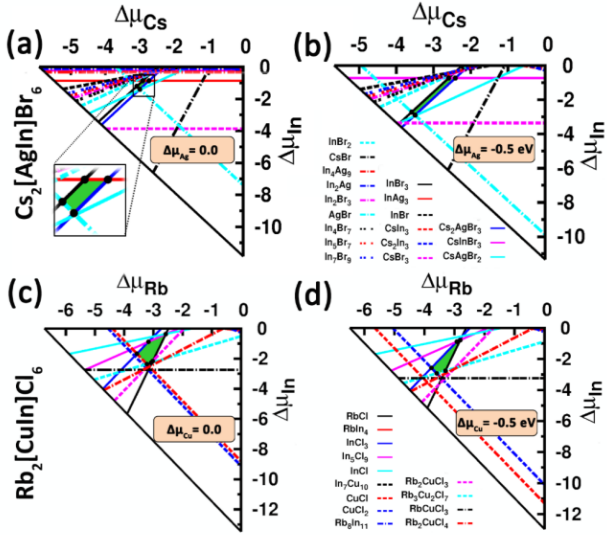

**Figure 3.** The phase stability diagram analysis results sliced at several Ag/Cu-varied growth conditions represented by $\Delta\mu_{Ag}/\Delta\mu_{Cu}$ (deviation of actual chemical potential of Ag/Cu from that of its metal phase) for $Cs_2[AgIn]Br_6$ (a and b) and $Rb_2[CuIn]Cl_6$ (c and d). The polygon region in green represents thermodynamic stable condition and each line corresponds to one competing phase.

***More precise thermodynamic stability analysis of the primarily stable solar CIHPs***: We perform thorough evaluation of thermodynamic stability via the phase stability diagram analysis[67,68] (as described in Supplementary Sec. I) for the 13 CIHPs passing the above initial screening. This takes fully into account all the decomposition channels into various combinations of the competing phases including all the existing binary, ternary, and quaternary compounds from the Inorganic Crystal Structure Database.[69] The results indicate that six CIHPs, *i.e.*, $Cs_2[AgIn]Cl_6$, $Cs_2[AgIn]Br_6$, $Rb_2[CuIn]Cl_6$, $Rb_2[CuIn]Br_6$, $Rb_2[AgIn]Cl_6$, and $Rb_2[AgIn]Br_6$ show thermodynamic stability evidenced by the visible polyhedron region in the three-dimensional space with chemical potential changes of constituted elements as variables. We note that the decomposition enthalpy with respect to the disproportionation channel into binary competing phases $\Delta H_{dec}$ (Fig. 2) of six thermodynamically stable CIHP compounds lie in the range of 11-116 meV/atom (positive values indicate stability with respect to disproportionation). The values are clearly more positive than those of $CH_3NH_3PbI_3$ as the latter was reported to have nearly zero or even negative $\Delta H_{dec}$.[3,70] This implies the much better materials stability with respect to disproportionation of the CIHP compounds.

Fig. 3 shows slices of the stable polyhedron region for $Cs_2[AgIn]Br_6$ (3a and 3b) and $Rb_2[CuIn]Cl_6$ (3c and 3d) taken at several Ag/Cu-varied growth conditions $\Delta\mu_{Ag}/\Delta\mu_{Cu}$ (deviation of actual chemical potential of Ag/Cu from that of elemental metal) (see more slices and results of other CIHPs in Supplementary Fig. S4-S9). The sliced polygon region stabilizing the CIHPs is marked in green and the lines surrounding it represent direct competing phases. One sees that both $Cs_2[AgIn]Br_6$ and $Rb_2[CuIn]Cl_6$ can be stabilized at the smaller magnitude of $\Delta\mu_{Ag}$ ($0 \geq \Delta\mu_{Ag} \geq -1.8$ eV, corresponding to Ag-rich conditions) and $\Delta\mu_{Cu}$ ($0 \geq \Delta\mu_{Cu} \geq -1.6$ eV, corresponding to Cu-rich conditions). Within the sliced plane with $\Delta\mu_{Rb}$ and $\Delta\mu_{In}$ as variables, the stable area of $Rb_2[CuIn]Cl_6$ is relatively large, especially at the growth condition of $\Delta\mu_{Cu} = 0$. This indicates its relative ease of being synthesized in terms of control of Rb and In contents. For $Cs_2AgInBr_6$, since the stable area is rather slim, careful control of elemental contents, *i.e.*, for both Cs and In at $\Delta\mu_{Ag} = 0$ and for Cs at $\Delta\mu_{Ag} = -0.5$ eV, are needed to grow high-quality samples by avoiding formation of secondary competing phases. Further discussions on experimental materials synthesis and four-element phase diagram for the stable CIHPs are provided below.

***Dynamic phonon stability of the thermodynamically stable CIHPs***: In addition to thermodynamic stability against decomposition into competing phases, phonon stability is another important quantity to characterize materials stability. Fig. 4a, 4b, and 4c show calculated harmonic phonon spectra (at 0 K) for $Cs_2[AgIn]Cl_6$ and $Cs_2[AgIn]Br_6$, as well as a Bi-based double-perovskites $Cs_2[AgBi]Cl_6$. One see that while $Cs_2[AgIn]Cl_6$ exhibits phonon stability evidenced by no imaginary modes, there are substantial imaginary optical branches (with frequencies up to –0.26 THz at the Γ point) in phonon spectrum of $Cs_2[AgIn]Br_6$. Surprisingly similar phonon instability occurs also to $Cs_2[AgBi]Cl_6$, the compound already synthesized in experiments.[5,6] This puzzling contradiction is resolved through calculating the room-temperature (300 K) phonon spectrum by including the anharmonic phonon-phonon interaction.[71] The calculated results are shown in Fig. 4d, 4e, and 4f for $Cs_2[AgIn]Cl_6$, $Cs_2[AgIn]Br_6$ and $Cs_2[AgBi]Cl_6$, respectively. Clearly the imaginary phonons of $Cs_2[AgBi]Cl_6$ are completely stabilized after taking into account the finite-temperature anharmonic effect. This is also the case for $Cs_2[AgIn]Br_6$. Therefore, the thermodynamically stable CIHPs we predicted have also the phonon stability at room temperature. Here phonon entropy at finite temperatures plays critical role in phonon spectrum renormalization. At low temperatures, involvement of atoms displacements with respect to the imaginary phonon modes may stabilize harmonic phonon spectra, leading to formation of the lower symmetric distorted perovskites containing tilted $Ag/Bi/InX_6$ octahedra.

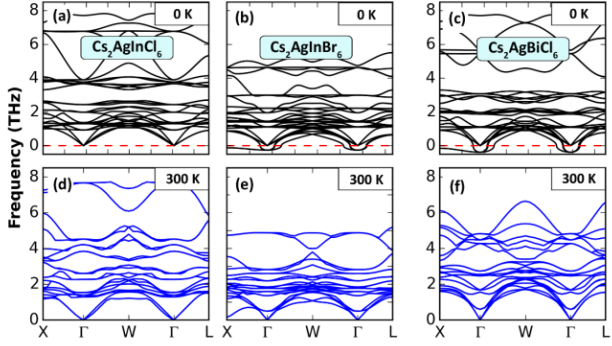

**Figure 4.** Calculated harmonic phonon spectra at 0 K (a, b, c) and room-temperature (300 K) phonon spectra taking into account anharmonic phonon-phonon interaction (d, e, f) for $Cs_2[AgIn]Cl_6$, $Cs_2[AgIn]Br_6$, and $Cs_2[AgBi]Cl_6$, respectively.

***Photovoltaic-related properties of the stable optimal solar CIHPs***: After identifying the six thermodynamically stable CIHPs with suitable direct band gaps, we then have a systemic exploration of their photovoltaic-related electronic and optoelectronic properties.

**(i) Carrier *effective masses*:** The Pb-based $APbX_3$ perovskites are known to have simultaneously low effective masses ($m^*$) of electrons and holes that contribute to their ambipolar conductivity and ultralong carrier diffusion length.[29,72] The calculated $m^*$ of both electrons and holes for six stable CIHPs are shown in Fig. 5e (corresponding data are listed in Supplementary Table S1). Since the VBM at Γ is doubly degenerated (Fig. 1d) and consists of two hole states (one is heavy and the other light, labeled as "hh" and "lh", respectively), the $m^*$ of both of them are calculated. This feature of combination of light electron and light-heavy hole effective masses, which resembles the cases of well-known solar materials of Cu[In,Ga]Se$_2$ chalcopyrites and III-V/II-VI semiconductors GaAs/CdTe, is expected to offer satisfactorily fast and balanced photon-induced carriers transport beneficial for high-efficiency solar energy conversion. One sees that all the CIHPs show low electron mass $m_e^*$ of 0.2-0.3$m_0$, much lower than the calculated value of $Cs_2[AgBi]Cl_6$ (0.92$m_0$) and even lower than that of $CH_3NH_3PbI_3$ (0.42$m_0$). For the hole states, the light hole mass $m_{lh}^*$ is as light as 0.3-0.4$m_0$, whereas the heavy hole mass $m_{hh}^*$ is approaching/above 2$m_0$.

With the obtained carriers effective masses we can roughly evaluate the exciton binding energy ($E_b$) by using the hydrogen-like Wannier-Mott exciton model.[3,52] The high-frequency limit of dielectric constant caused by electronic polarization, are calculated for this purpose; the resulted $E_b$ describes the exciton generated immediately after photon excitation (without lattice polarization process involved). The results for six stable CIHPs are summarized in Supplementary Table S1. Except for $Cs_2[AgIn]Cl_6$ and $Rb_2[AgIn]Cl_6$ with the large band gaps (~2.5 eV), other compounds generally show moderate $E_b$ values below or mildly above 100 meV. The values are comparable to the calculated ones in the same approach for Pb-based hybrid iodide perovskites.[3] More advanced Bethe–Salpeter equation calculation of exciton would provide more definitive values but beyond the scope of the current study.

**(ii) *Electronic band structure*:** As shown in Fig 1d, the band structure of $Cs_2[AgIn]Cl_6$, a stable CIHP shows clearly *direct gap* opened at the Γ point of the Brillouin zone. This is distinct from the case of Bi-based double-perovskites $Cs_2[AgBi]Cl_6$ (Fig. 1b) having an indirect band gap formed between the VBM at X and CBM at L (nearly degenerate with the Γ state). The reason for the different band-structure features is as follows. In $Cs_2[Ag^+In^{3+}]Cl_6$, the cation complex $[Ag^+ + In^{3+}]$ consists of a bare ion $In^{3+}$ (in $s^0p^0$ electron configuration) lacking any valence electrons and thus contributing its empty $s$ and $p$ orbitals to the conduction band, whereas the $Ag^+$ ion (in $4d^{10}$ configuration) is the sole cation that participates in forming the valence bands. This cation complex thus does not have the cationic states hybridization that is present between the *occupied* valence states of Ag($d$) and Bi($s$), which is the ultimate cause of indirect band gap in $Cs_2[AgBi]Cl_6$.[7,52] As unambiguously demonstrated in the orbital-projected band edges (Fig. 1d) and projected density of states (Fig. 5a for $Cs_2AgInBr_6$), the valence bands of $Cs_2[AgIn]Cl_6$ is cleanly formed by the hybridization between cationic Ag($d$) and anionic Cl($p$) orbitals, and thus forms its maximum at Γ. Therefore, the band gap becomes direct. We note that the top valence band along the Γ-X direction is nearly degenerate in energy with the VBM at Γ, resulting in a rather flat band with extremely heavy $m^*$. The Γ-X direction corresponds to the direction between the nearest-neighboring Ag and In ions in real space. This band originates from the Ag ($d_{x^2-y^2}$) orbital that weakly couples with the Cl($p$) orbital (as seen from its orbital-projection in Fig. 1d).

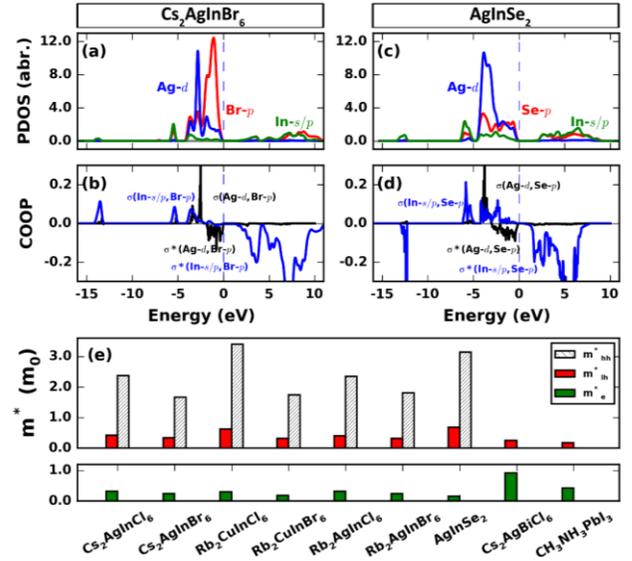

**Figure 5**. Atomic orbital projected density of states, PDOS (a, c) and crystal orbital overlap population (COOP) for bonding-type analysis (b, d) for the CIHP $Cs_2[AgIn]Br_6$ and the chalcopyrite $AgInSe_2$, respectively. (e) Calculated carrier effective masses ($m^*$) of six stable CIHPs, compared with the results of $AgInSe_2$, $CH_3NH_3PbI_3$, and $Cs_2[AgBi]Cl_6$. The $m^*$ values are calculated in terms of second derivative of the HSE-functional-derived band dispersion curve (along the Γ-W direction for CIHPs and $Cs_2[AgBi]Cl_6$, the Γ-X direction for $AgInSe_2$, and the Γ-W direction for $CH_3NH_3PbI_3$).

Fig. 5a and 5b show the projected density of states and crystal orbital overlap population (COOP) for bonding-type analysis[73] of $Cs_2[AgIn]Br_6$, in comparison with the results of chalcopyrite $AgInSe_2$ (Fig. 5c and 5d). The positive COOP means bonding states, whereas negative represents anti-bonding ones. Clearly, we see that the upper valence bands of $Cs_2[AgIn]Br_6$ are composed of anti-bonding states of the Ag(*d*) and Br(*p*) orbitals. The conduction bands are predominated by the In(*s/p*) orbitals that couple with the Br(*p*) orbital in the anti-bonding manner as well. These electronic structure features, in close similarity with those of $AgInSe_2$, contribute to a rather strong light absorption in proximity to band gap (as shown below) and is expected to offer the superior defect tolerant feature[11] beneficial for efficient carrier extraction and transport.

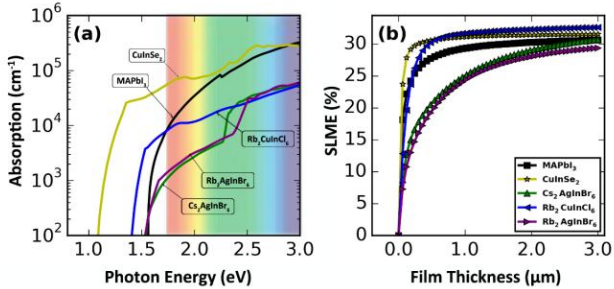

**Figure 6**. (a) Calculated photon absorption spectra of $Cs_2[AgIn]Br_6$, $Rb_2[AgIn]Br_6$, and $Rb_2[CuIn]Cl_6$, compared with the cases of chalcopyrite $CuInSe_2$ and Pb-based hybrid halide perovskite $CH_3NH_3PbI_3$. (b) Calculated "spectroscopic limited maximum efficiency (SLME)" based on the improved Shockley-Queisser model.

**(iii) *Optical absorption*:** Fig. 6a shows calculated light absorption spectra of three optimal CIHPs with more ideal solar band gaps, $Cs_2[AgIn]Br_6$ (1.50 eV), $Rb_2[CuIn]Cl_6$ (1.36 eV), and $Rb_2[AgIn]Br_6$ (1.46 eV), compared with those of chalcopyrite $CuInSe_2$ and $CH_3NH_3PbI_3$. As expected from the direct-gap nature, $Cs_2[AgIn]Br_6$, $Rb_2[CuIn]Cl_6$, and $Rb_2[AgIn]Br_6$ show rather strong absorption edges. The absorption near threshold is contributed by the optical transition channels from the [Ag,Cu](*d*)/[Br,Cl](*p*) orbitals dominating valence bands to the In(*s/p*) orbitals of conduction bands. Though with the same direct-gap features, the band-edge absorptions of $Cs_2[AgIn]Br_6$ and $Rb_2[AgIn]Br_6$ are lower than that of chalcopyrite $CuInSe_2$ (and also $CH_3NH_3PbI_3$). At about 2.3 and 2.4 eV, $Cs_2[AgIn]Br_6$ and $Rb_2[AgIn]Br_6$ exhibit small-magnitude fast increase in absorption. Different from the two Ag-based compounds, $Rb_2[CuIn]Cl_6$ shows the stronger band-edge absorption and no emergence of the second small-magnitude fast increase. Direct comparisons of band structure and joint density of states (JDOS) between $Cs_2[AgIn]Br_6$ and $Rb_2[CuIn]Cl_6$ (Supplementary Fig. S10) show that $Rb_2[CuIn]Cl_6$ has the higher JDOS in proximity to the band gap threshold than that of $Cs_2[AgIn]Br_6$. This originates from its less dispersive top valence bands, possibly due to the weaker antibonding hybridization between Cu-*d* and Cl-*p* orbitals. The implication of this is that the absorption intensity near band gap of $A_2BCX_6$ CIHPs could be enhanced by deliberate chemical composition engineering, for instance through doing alloying on the B/C and X sites, or optimizing the A-site cations.

**(iv) *Solar cell efficiency*:** Fig. 6b shows the calculated "spectroscopic limited maximum efficiency (SLME)"[65] with optical absorption spectrum and thickness of thin-film absorber as inputs. Compared with $CuInSe_2$ and $CH_3NH_3PbI_3$, $Cs_2[AgIn]Br_6$ and $Rb_2[AgIn]Br_6$ show a smoothly gradual increasing SLME with the increasing film thickness. This ascribes to their lower absorption intensity in proximity to threshold as mentioned. This, however, does not impede $Cs_2[AgIn]Br_6$ and $Rb_2[AgIn]Br_6$ being promising good solar absorbers, since at the film thickness of 1 μm, their SLME values have been above 20%. At the 2 μm thickness of thin films, their SLME values reach about 28%, becoming comparable to the values of $CuInSe_2$ (31.5%) and $CH_3NH_3PbI_3$ (30%). Turning to $Rb_2[CuIn]Cl_6$, it exhibits a sharp increment of SLME, quite similar to $CuInSe_2$ and $CH_3NH_3PbI_3$. At the film thickness of 1 μm its SLME reaches 31.7%, even surpassing the values of $CuInSe_2$ and $CH_3NH_3PbI_3$. This is attributed to its calculated band gap (1.36 eV) in close proximity to the optimal one determined by the Shockley–Queisser limit (1.34 eV).

Finally for good photovoltaic materials, the existing defects need to be shallow, rather than deep mid-gap states to guarantee bipolar conductivity and avoid detrimental carrier trapping centers. A full assessment of defect physics would require the calculation of the formation energies and transition levels of all possible native defects, which is outside the scope of the current first paper introducing this new group of CIHP compounds. We note however that the bulk electronic structure demonstrated here provides some clues to the nature of the defects. For instance in $Cs_2[AgIn]Br_6$, the anti-bonding (between Ag-*d* and Br-*p* orbitals) nature of the upper valence bands is expected to make $V_{Ag}$ *p*-type shallow defect, resembling $V_{Cu}$ of Cu[In,Ga]Se$_2$.[54,55] Considering its rather strong ionic bonding, $Cs_i$ and $V_{Br}$ would be low-energy *n*-type defects with shallow transition levels, similar to the case of Pb-based hybrid perovskites.[30-32]

***Theoretical comments on materials synthesis:*** To synthesize an unknown material with new stoichiometry, one has to control synthesis condition (such as temperature, pressure, chemical potential of compositions, *etc.*) to avoid formation of unwanted competing phases. For the six proposed optimal CIHP $A_2BCX_6$ compounds, we provide the tetrahedral phase diagram of quaternary A-B-C-X system (as shown in Supplementary Fig. S11). All the known existing (binary and ternary) phases (considered by the phase stability diagram analysis in Fig. 3 and Supplementary Figs. S4-S9) are mapped into the tetrahedron with elemental compositions of A, B, C, and X as vertexes. The directly competing phases, which critically control the CIHP compound stability and correspond to the lines surrounding the green stable polygon region in Fig. 3, are shown in blue. Taking $Cs_2[AgIn]Br_6$ as example, the thermodynamic stability against disproportionation into competing phases (Figs. 3a and 3b) is best met under Ag-rich condition (with $\Delta\mu_{Ag}$ being close to 0). This means the content of Ag-containing precursor needs to be abundant during synthesis. The directly competing phases are three binary compounds of $InBr_3$, $InAg_3$, $AgBr$, and three ternary ones of $Cs_2AgBr_3$, $CsAgBr_2$, $CsInBr_3$ (Supplementary Fig. S11f). Avoiding the reactions to form these competing phases through controlling synthesis condition would facilitate experimental synthesis of $Cs_2[AgIn]Br_6$.

## 4. Conclusions

We propose to overcome the limitations of Pb-based halide perovskites $APbX_3$ and Bi-based double perovskites $A_2[AgBi]X_6$ not by trying empirically different substitutions that can be gleaned from databases, but rather by following the pertinent "design principles" that emerge from understanding of electronic structure underpinning of past successes and failures. This led us to design a class of stable, Pb-free double perovskites $A_2[BC]X_6$ where 2Pb in $APbX_3$ is transmuted by the element pair [B, C] that made Cu-based chalcopyrites the leading thin-film photovoltaic absorber: [Cu/Ag, Ga/In]. This constitutes a new group of **CuIn–based Halide Perovskite** (CIHP), which combines the paradigm of the superior Cu($d$)Se($p$)→In($s/p$) valence-to-conduction-band absorption channel underlying chalcopyrites, with the halide perovskite structure providing $A^+$-$[BX_6/CX_6]^-$ ionic bonding and at the same time covalent-like not big band gaps. Meanwhile, switching from C = $Bi^{3+}$ in $ns^2$ electron configuration in $A_2[AgBi]X_6$ to C = $In^{3+}$ in $ns^0$ configuration in $A_2[AgIn]X_6$ eliminates the electronic symmetry reduction due to the orbital mismatch between Ag($d$) and Bi($s$) which caused indirect band gap in $A_2[AgBi]X_6$. This enables the creation of direct-gap double perovskites.

First-principles calculations are employed to study materials stability and optoelectronic properties of a series of $A_2[BC]X_6$ CIHPs with $A$ = K, Rb, Cs; $B$ = Cu, Ag; C = Ga, In; X = Cl, Br, I. We find the family of CIHPs shows desired direct band gaps in a wide tunable range from 0 to about 2.5 eV, rather strong absorption spectra near threshold, as well as combination of light electron and heavy-light hole effective masses. The superior properties for photovoltaics originate from the closed Cu/Ag($d^{10}$) shell which dominates valence bands by anti-bondingly hybridizing with X($p$) orbitals, resembling that of chalcopyrites. We identify via materials screening six CIHPs that are chemically stable with respect to disproportionation and dynamically stable evidenced by absence of imaginary phonons. These include $Cs_2[AgIn]Cl_6$, $Cs_2[AgIn]Br_6$, $Rb_2[AgIn]Cl_6$, $Rb_2[AgIn]Br_6$, $Rb_2[CuIn]Cl_6$, and $Rb_2[CuIn]Br_6$ with the calculated direct band gaps of 2.52, 1.50, 2.50, 1.46, 1.36 and 0.63 eV, respectively. Particularly $Cs_2[AgIn]Br_6$, $Rb_2[AgIn]Br_6$, and $Rb_2[CuIn]Cl_6$ with the direct gaps close to the optimal value of 1.34 eV show large "spectroscopic limited maximum efficiency" of nearly 30% for films thickness of ~2 μm — comparable to the photovoltaic performance of chalcopyrites and $CH_3NH_3PbI_3$. Solid solution among these table compounds with different gap values *e.g.*, $Cs_2[AgIn](Br,Cl)_6$ and $Rb_2[(Cu,Ag)In]Br_6$, would offer further opportunity for band gap engineering, suggesting the possibility of using them in tandem solar cells. The predicted stabilities, low environment effect and superior photovoltaic performance suggest that these materials would be the alternative novel ones for the solar absorbers. Our findings enrich the family of photovoltaic halide perovskites and experimental efforts in synthesizing our predicted materials are called for.

During revision of the manuscript upon editorial request, we become aware of a related joint experiment-theory study by Volonakis *et al.*[74] on the CIHP compounds, which has synthesized $Cs_2[AgIn]Cl_6$ showing direct band gap.

## Supporting Information
More detailed computational procedures, energy order of the different structural configurations for $A_2BCX_6$ perovskites, comparison between calculated and experimental data of the classical chalcopyrites, additional supporting data on materials stability analysis, explicit calculated lattice constants and band gaps of all the candidate $A_2BCX_6$ used for materials screening.

## Corresponding Author
*lijun_zhang@jlu.edu.cn

## Acknowledgement
The authors acknowledge funding support from the Recruitment Program of Global Youth Experts in China, National Key Research and Development Program of China (under Grants No. 2016YFB0201204), National Natural Science Foundation of China (under Grant No. 11404131 and 11674121). The work at CU was supported by the EERE Sun Shot (on the stability of materials) and fundamental research on the new material discovery was supported by Office of Science, Basic Energy Science, MSE Division under DE-SC0010467. Part of calculations was performed in the high-performance computing center of Jilin University and on TianHe-1(A) of National Supercomputer Center in Tianjin.

## References
(1) Noel, N. K.; Stranks, S. D.; Abate, A.; Wehrenfennig, C.; Guarnera, S.; Haghighirad, A.-A.; Sadhanala, A.; Eperon, G. E.; Pathak, S. K.; Johnston, M. B.; Petrozza, A.; Herz, L. M.; Snaith, H. J. *Energy Environ. Sci.* **2014**, *7*, 3061–3068.
(2) Hao, F.; Stoumpos, C. C.; Cao, D. H.; Chang, R. P. H.; Kanatzidis, M. G. *Nat. Photonics* **2014**, *8*, 489–494.
(3) Yang, D.; Lv, J.; Zhao, X.; Xu, Q.; Fu, Y.; Zhan, Y.; Zunger, A.; Zhang, L. *Chem. Mater.* **2016**, *29*, 524–538.
(4) Deng, Z.; Wei, F.; Sun, S.; Kieslich, G.; Cheetham, A. K.; Bristowe, P. D. *J Mater Chem A* **2016**, *4*, 12025–12029.
(5) Slavney, A. H.; Hu, T.; Lindenberg, A. M.; Karunadasa, H. I. *J. Am. Chem. Soc.* **2016**, *138*, 2138–2141.
(6) McClure, E. T.; Ball, M. R.; Windl, W.; Woodward, P. M. *Chem. Mater.* **2016**, *28*, 1348–1354.
(7) Savory, C. N.; Walsh, A.; Scanlon, D. O. *ACS Energy Lett.* **2016**, *1*, 949–955.
(8) Volonakis, G.; Filip, M. R.; Haghighirad, A. A.; Sakai, N.; Wenger, B.; Snaith, H. J.; Giustino, F. *J. Phys. Chem. Lett.* **2016**, *7*, 1254–1259.
(9) Park, B.-W.; Philippe, B.; Zhang, X.; Rensmo, H.; Boschloo, G.; Johansson, E. M. J. *Adv. Mater.* **2015**, *27*, 6806–6813.
(10) Chirilă, A.; Buecheler, S.; Pianezzi, F.; Bloesch, P.; Gretener, C.; Uhl, A. R.; Fella, C.; Kranz, L.; Perrenoud, J.; Seyrling, S.; Verma, R.; Nishiwaki, S.; Romanyuk, Y. E.; Bilger, G.; Tiwari, A. N. *Nat. Mater.* **2011**, *10*, 857–861.
(11) Zhang, S. B.; Wei, S.-H.; Zunger, A.; Katayama-Yoshida, H. *Phys. Rev. B* **1998**, *57*, 9642–9656.
(12) Persson, C.; Zunger, A. *Phys. Rev. Lett.* **2003**, *91*, 266401.
(13) Filip, M. R.; Hillman, S.; Haghighirad, A. A.; Snaith, H. J.; Giustino, F. *J. Phys. Chem. Lett.* **2016**, *7*, 2579–2585.
(14) Jackson, P.; Hariskos, D.; Wuerz, R.; Wischmann, W.; Powalla, M. *Phys. Status Solidi RRL – Rapid Res. Lett.* **2014**, *8*, 219–222.
(15) Guillemoles, J.-F.; Kronik, L.; Cahen, D.; Rau, U.; Jasenek, A.; Schock, H.-W. *J. Phys. Chem. B* **2000**, *104*, 4849–4862.
(16) Mitzi, D. B. In *Progress in Inorganic Chemistry*; Karlin, K. D., Ed.; John Wiley & Sons, Inc., New York, 1999.
(17) Kojima, A.; Teshima, K.; Shirai, Y.; Miyasaka, T. *J. Am. Chem. Soc.* **2009**, *131*, 6050–6051.
(18) Kim, H.-S.; Lee, C.-R.; Im, J.-H.; Lee, K.-B.; Moehl, T.; Marchioro, A.; Moon, S.-J.; Humphry-Baker, R.; Yum, J.-H.; Moser, J. E.; Grätzel, M.; Park, N.-G. *Sci. Rep.* **2012**, *2*, 591.
(19) Noh, J. H.; Im, S. H.; Heo, J. H.; Mandal, T. N.; Seok, S. I. *Nano Lett.* **2013**, *13*, 1764–1769.
(20) Burschka, J.; Pellet, N.; Moon, S.-J.; Humphry-Baker, R.; Gao, P.; Nazeeruddin, M. K.; Grätzel, M. *Nature* **2013**, *499*, 316–319.

(21) Jeon, N. J.; Noh, J. H.; Kim, Y. C.; Yang, W. S.; Ryu, S.; Seok, S. I. *Nat. Mater.* **2014**, *13*, 897–903.
(22) Im, J.-H.; Jang, I.-H.; Pellet, N.; Grätzel, M.; Park, N.-G. *Nat. Nanotechnol.* **2014**, *9*, 927–932.
(23) Yang, W. S.; Noh, J. H.; Jeon, N. J.; Kim, Y. C.; Ryu, S.; Seo, J.; Seok, S. I. *Science* **2015**, *348*, 1234–1237.
(24) Chen, W.; Wu, Y.; Yue, Y.; Liu, J.; Zhang, W.; Yang, X.; Chen, H.; Bi, E.; Ashraful, I.; Grätzel, M.; Han, L. *Science* **2015**, *350*, 944–948.
(25) NREL, http://www.nrel.gov/ncpv/images/efficiency_chart.jpg (accessed June 6th, 2016).
(26) Yin, W.-J.; Shi, T.; Yan, Y. *Adv. Mater.* **2014**, *26*, 4653–4658.
(27) D'Innocenzo, V.; Grancini, G.; Alcocer, M. J. P.; Kandada, A. R. S.; Stranks, S. D.; Lee, M. M.; Lanzani, G.; Snaith, H. J.; Petrozza, A. *Nat. Commun.* **2014**, *5*, 3586.
(28) Miyata, A.; Mitioglu, A.; Plochocka, P.; Portugall, O.; Wang, J. T.-W.; Stranks, S. D.; Snaith, H. J.; Nicholas, R. J. *Nat. Phys.* **2015**, *11*, 582–587.
(29) Stranks, S. D.; Eperon, G. E.; Grancini, G.; Menelaou, C.; Alcocer, M. J. P.; Leijtens, T.; Herz, L. M.; Petrozza, A.; Snaith, H. J. *Science* **2013**, *342*, 341–344.
(30) Hutter, E. M.; Eperon, G. E.; Stranks, S. D.; Savenije, T. J. *J. Phys. Chem. Lett.* **2015**, *6*, 3082–3090.
(31) Wu, X.; Trinh, M. T.; Niesner, D.; Zhu, H.; Norman, Z.; Owen, J. S.; Yaffe, O.; Kudisch, B. J.; Zhu, X.-Y. *J. Am. Chem. Soc.* **2015**, *137*, 2089–2096.
(32) Yin, W.-J.; Shi, T.; Yan, Y. *Appl. Phys. Lett.* **2014**, *104*, 63903.
(33) Philippe, B.; Park, B.-W.; Lindblad, R.; Oscarsson, J.; Ahmadi, S.; Johansson, E. M. J.; Rensmo, H. *Chem. Mater.* **2015**, *27*, 1720–1731.
(34) Ye, F.; Yang, W.; Luo, D.; Zhu, R.; Gong, Q. *J. Semicond.* **2017**, *38*, 11003.
(35) Shirayama M.; Kato M.; Miyadera T.; Sugita S.; Fujiseki T.; Hara S.; Kadowaki H.; Murata D.; Chikamatsu M. and Fujiwara H. *J. Appl. Phys.* **2016**, *119*, 115501.
(36) Bryant, D.; Aristidou, N.; Pont, S.; Sanchez-Molina, I.; Chotchunangatchaval, T.; Wheeler, S.; Durrant, J. R.; Haque, S. A. *Energy Environ. Sci.* **2016**, *9*, 1655–1660.
(37) Akbulatov, A. F.; Luchkin, S. Y.; Frolova, L. A.; Dremova, N. N.; Gerasimov, K. L.; Zhidkov, I. S.; Anokhin, D. V.; Kurmaev, E. Z.; Stevenson, K. J.; Troshin, P. A. *J. Phys. Chem. Lett.* **2017**, *8*, 1211–1218.
(38) Leguy, A. M. A.; Hu, Y.; Campoy-Quiles, M.; Alonso, M. I.; Weber, O. J.; Azarhoosh, P.; van Schilfgaarde, M.; Weller, M. T.; Bein, T.; Nelson, J.; Docampo, P.; Barnes, P. R. F. *Chem. Mater.* **2015**, *27*, 3397–3407.
(39) Slavney, A. H.; Smaha, R. W.; Smith, I. C.; Jaffe, A.; Umeyama, D.; Karunadasa, H. I. *Inorg. Chem.* **2016**, *56*, 46–55.
(40) Conings, B.; Drijkoningen, J.; Gauquelin, N.; Babayigit, A.; D'Haen, J.; D'Olieslaeger, L.; Ethirajan, A.; Verbeeck, J.; Manca, J.; Mosconi, E.; Angelis, F. D.; Boyen, H.-G. *Adv. Energy Mater.* **2015**, *5*, 1500477.
(41) Jeon, N. J.; Noh, J. H.; Yang, W. S.; Kim, Y. C.; Ryu, S.; Seo, J.; Seok, S. I. *Nature* **2015**, *517*, 476–480.
(42) Filip, M. R.; Giustino, F. *J. Phys. Chem. C* **2016**, *120*, 166–173.
(43) McMeekin, D. P.; Sadoughi, G.; Rehman, W.; Eperon, G. E.; Saliba, M.; Hörantner, M. T.; Haghighirad, A.; Sakai, N.; Korte, L.; Rech, B.; Johnston, M. B.; Herz, L. M.; Snaith, H. J. *Science* **2016**, *351*, 151–155.
(44) Hao, F.; Stoumpos, C. C.; Chang, R. P. H.; Kanatzidis, M. G. *J. Am. Chem. Soc.* **2014**, *136*, 8094–8099.
(45) Eperon, G. E.; Leijtens, T.; Bush, K. A.; Prasanna, R.; Green, T.; Wang, J. T.-W.; McMeekin, D. P.; Volonakis, G.; Milot, R. L.; May, R.; Palmstrom, A.; Slotcavage, D. J.; Belisle, R. A.; Patel, J. B.; Parrott, E. S.; Sutton, R. J.; Ma, W.; Moghadam, F.; Conings, B.; Babayigit, A.; Boyen, H.-G.; Bent, S.; Giustino, F.; Herz, L. M.; Johnston, M. B.; McGehee, M. D.; Snaith, H. J. *Science* **2016**, *354*, 861–865.
(46) Li, Z.; Yang, M.; Park, J.-S.; Wei, S.-H.; Berry, J. J.; Zhu, K. *Chem. Mater.* **2016**, *28*, 284–292.
(47) Swarnkar, A.; Marshall, A. R.; Sanehira, E. M.; Chernomordik, B. D.; Moore, D. T.; Christians, J. A.; Chakrabarti, T.; Luther, J. M. *Science* **2016**, *354*, 92–95.
(48) Saliba, M.; Matsui, T.; Seo, J.-Y.; Domanski, K.; Correa-Baena, J.-P.; Nazeeruddin, M. K.; Zakeeruddin, S. M.; Tress, W.; Abate, A.; Hagfeldt, A.; Grätzel, M. *Energy Environ. Sci.* **2016**, *9*, 1989–1997.
(49) Saliba, M.; Matsui, T.; Domanski, K.; Seo, J.-Y.; Ummadisingu, A.; Zakeeruddin, S. M.; Correa-Baena, J.-P.; Tress, W. R.; Abate, A.; Hagfeldt, A.; Grätzel, M. *Science* **2016**, *354*, 206–209.
(50) Biswas, K.; Lany, S.; Zunger, A. *Appl. Phys. Lett.* **2010**, *96*, 201902.
(51) Jacobsson, T. J.; Pazoki, M.; Hagfeldt, A.; Edvinsson, T. *J. Phys. Chem. C* **2015**, *119*, 25673–25683.
(52) Zhao, X.; Yang, J.; Fu, Y.; Yang, D.; Xu, Q.; Yu, L.; Wei, S.-H.; Zhang, L. *J. Am. Chem. Soc.* **2017**, *139*, 2630–2638.
(53) Noufi, R.; Axton, R.; Herrington, C.; Deb, S. K. *Appl. Phys. Lett.* **1984**, *45*, 668–670.
(54) Wei, S.-H.; Zhang, S. B.; Zunger, A. *Appl. Phys. Lett.* **1998**, *72*, 3199–3201.
(55) Jaffe, J. E.; Zunger, A. *Phys. Rev. B* **2001**, *64*, 241304.
(56) Yan, Y.; Noufi, R.; Al-Jassim, M. M. *Phys. Rev. Lett.* **2006**, *96*, 205501.
(57) Abou-Ras, D.; Schaffer, B.; Schaffer, M.; Schmidt, S. S.; Caballero, R.; Unold, T. *Phys. Rev. Lett.* **2012**, *108*, 75502.
(58) Schmidt, S. S.; Abou-Ras, D.; Sadewasser, S.; Yin, W.; Feng, C.; Yan, Y. *Phys. Rev. Lett.* **2012**, *109*, 95506.
(59) Kresse, G.; Furthmüller, J. *Comput. Mater. Sci.* **1996**, *6*, 15–50.
(60) Kresse, G.; Joubert, D. *Phys. Rev. B* **1999**, *59*, 1758–1775.
(61) Perdew, J. P.; Burke, K.; Ernzerhof, M. *Phys. Rev. Lett.* **1996**, *77*, 3865–3868.
(62) Krukau, A. V.; Vydrov, O. A.; Izmaylov, A. F.; Scuseria, G. E. *J. Chem. Phys.* **2006**, *125*, 224106.
(63) Togo, A.; Tanaka, I. *Scr. Mater.* **2015**, *108*, 1–5.
(64) Souvatzis, P.; Eriksson, O.; Katsnelson, M. I.; Rudin, S. P. *Phys. Rev. Lett.* **2008**, *100*, 95901.
(65) Yu, L.; Zunger, A. *Phys. Rev. Lett.* **2012**, *108*, 68701.
(66) Han, Y.; Meyer, S.; Dkhissi, Y.; Weber, K.; Pringle, J. M.; Bach, U.; Spiccia, L.; Cheng, Y.-B. *J. Mater. Chem. A* **2015**, *3*, 8139–8147.
(67) Li, Y.; Singh, D. J.; Du, M.-H.; Xu, Q.; Zhang, L.; Zheng, W.; Ma, Y. *J. Mater. Chem. C* **2016**, *4*, 4592–4599.
(68) Persson, C.; Zhao, Y.-J.; Lany, S.; Zunger, A. *Phys. Rev. B* **2005**, *72*, 35211.
(69) Belsky, A.; Hellenbrandt, M.; Karen, V. L.; Luksch, P. *Acta Crystallogr. B* **2002**, *58*, 364–369.
(70) Zhang, Y.-Y.; Chen, S.; Xu, P.; Xiang, H.; Gong, X.-G.; Walsh, A.; Wei, S.-H. **2015**, *ArXiv:150601301*.
(71) Souvatzis, P.; Eriksson, O.; Katsnelson, M. I.; Rudin, S. P. *Comput. Mater. Sci.* **2009**, *44*, 888–894.
(72) Xing, G.; Mathews, N.; Sun, S.; Lim, S. S.; Lam, Y. M.; Gratzel, M.; Mhaisalkar, S.; Sum, T. C. *Science* **2013**, *342*, 344–347.
(73) Hoffman, R. *Solids and Surfaces: A Chemist's View of Bonding in Extended Structures*; VCH Publishers, Inc., New York, 1988.
(74) Volonakis, G.; Haghighirad, A. A.; Milot, R. L.; Sio, W. H.; Filip, M. R.; Wenger, B.; Johnston, M. B.; Herz, L. M.; Snaith, H. J.; Giustino, F. *J. Phys. Chem. Lett.* **2017**, *8*, 772–778.

# Supporting Information

## Cu-In Halide Perovskite solar absorbers

Xin-Gang Zhao[1], Dongwen Yang[1], Yuanhui Sun[1], Tianshu Li[1], Lijun Zhang[1],*, Liping Yu[2], and Alex Zunger[3]

[1]State Key Laboratory of Superhard Materials, Key Laboratory of Automobile Materials of MOE, and College of Materials Science and Engineering, Jilin University, Changchun 130012, China

[2]Department of Physics, Temple University, Philadelphia, PA 19122, USA

[3]University of Colorado, Renewable and Sustainable Energy Institute, Boulder, Colorado 80309 USA

**Corresponding Author**
* lijun_zhang@jlu.edu.cn

I. Detailed Computational Procedures

*Structure optimization:* All the candidate $A_2[BC]X_6$ CIHPs in the cubic double-perovskite *Fm-3m* structure are optimized theoretically via total energy minimization with the conjugate-gradient algorithm. Both lattice parameters and internal atomic coordinates are fully relaxed. We used the high enough kinetic energy cutoffs for the plane-wave basis sets, *i.e.*, K/Rb/Cs: 337/286/286 eV; Cu/Ag: 355/325 eV; Ga/In: 175/125 eV; Cl/Br/I: 364/281/228 eV, to eliminate the potential Pulay stress error during crystalline cell optimization. The *k*-points meshes with grid spacing of less than $2\pi \times 0.10$ Å$^{-1}$ are used for electronic Brillouin zone integration. The convergence threshold for the residual forces on atoms is set to 0.0002 eV/Å.

*Band Gap:* As known, DFT calculations usually seriously underestimate (by ~50-100%) the band gaps of most of semiconducting materials. To remedy this problem, we employ the Heyd-Scuseria-Ernzerhof (HSE) hybrid functional approach[1] to reduce the self-interaction error and approach real gap values. The standard 25% exact Fock exchange is included. The HSE functional is used both for structural optimization and for evaluating the band gap at the optimized geometry. After obtaining the reliable HSE band gaps, the band structure, density of states, and absorption spectrum from the DFT-PBE calculations are corrected by the scissor operator to match the HSE gap values.

*Phase stability diagram analysis:* To guarantee a stable $A_2[BC]X_6$ CIHP in materials growth, thermodynamic equilibrium condition requires that the following three relations need to be satisfied.[2,3]

$$2\Delta\mu_A + \Delta\mu_B + \Delta\mu_C + 6\Delta\mu_X = \Delta H_f(A_2BCX_6), \quad (1)$$

$$\Delta\mu_i \leq 0 \,(i = A, B, C, X), \quad (2)$$

$$h_j\Delta\mu_A + k_j\Delta\mu_B + n_j\Delta\mu_C + m_j\Delta\mu_X \leq \Delta H_f(A_{h_j}B_{k_j}C_{n_j}X_{m_j}), j = 1 \cdots Z, \quad (3)$$

where $\Delta\mu_i = \mu_i - \mu_i^0$ is deviation of the chemical potential of atomic specie *i* during growth ($\mu_i$) from that of its solidified or gas phase ($\mu_i^0$), $\Delta H_f$ is heat of formation, and $A_{h_j}B_{k_j}C_{n_j}X_{m_j}$ represents all the existing competing phases (with the total number of *Z*). Eq. (1) is for thermodynamic equilibrium, eq. (2) is to prevent atomic species from precipitating to elemental phases, and eq. (3) is to avoid formation of any secondary competing phase. Eq. (1) determines only three $\Delta\mu_i$ are independent. Solutions to this group of equations, *i.e.*, the ranges of $\Delta\mu_i$ that stabilize the $A_2[BC]X_6$ CIHP, are bound in a polyhedron in the three-dimensional space with three $\Delta\mu_i$ as variables.

*Phonon spectrum:* To evaluate dynamical phonon stability, we calculate harmonic phonon spectrum (at 0 K) and room-temperature (300 K) phonon spectrum with inclusion of phonon-phonon interactions (anharmonic effects). The harmonic phonon spectrum is calculated from second-order interatomic force constants obtained by using the real-space finite-difference approach implemented in Phonopy code.[4] The 2×2×2 supercell (of the primitive cell of the double-perovskite structure) accompanying with the *k*-point mesh with grid spacing of $2\pi \times 0.03$ Å$^{-1}$ is used for these calculations. The room-temperature phonon spectrum is obtained by taking into account anharmonic phonon-phonon interaction with a self-consistent *ab initio* lattice dynamical (SCAILD) method.[5] This is done via calculating the phonon frequencies renormalization induced by phonon entropy, *i.e.*, the geometric disorder introduced by several frozen phonons simultaneously presenting in the simulated supercell. The SCAILD method alternates between creating atomic displacements in terms of phonon modes and evaluating phonon frequencies from calculated forces acting on the displaced atoms. The self-consistent cycle was terminated when the difference in the system free energy between two consecutive iterations is less than 1 meV. Calculations are performed at constant volume with thermal expansion effect ignored.

*Absorption spectrum:* The photon energy ($\omega$) dependent absorption coefficient $\alpha(\omega)$ is calculated from real/imaginary parts of dielectric function[$\varepsilon_1(\omega)/\varepsilon_2(\omega)$]. The $\varepsilon_2(\omega)$ is calculated in the random phase approximation,[6] and $\varepsilon_1(\omega)$ is evaluated from $\varepsilon_2(\omega)$ via the Kramers-Kronig relation. The dense *k*-point meshes with grid spacing of less than $2\pi \times 0.015$ Å$^{-1}$ is used for calculating ground-state band structure to guarantee $\varepsilon_2(\omega)$ converged. The twice of the number of occupied valence bands is used for calculating empty conduction band states.

*Maximum solar cell efficiency:* The maximum solar cell efficiency is simulated through calculating spectroscopic limited maximum efficiency (SLME) based on the improved Shockley-Queisser model. The detailed calculation procedure was described elsewhere.[7,8] It takes into account the effects of key intrinsic materials properties such as band gap, shape of absorption spectra, and material-dependent nonradiative recombination losses, on the photovoltaic efficiency. The simulation is performed under the standard AM1.5G solar spectrum at room temperature.

## II. Goldschmidt's empirical rule on formability of halide perovskites

To approximately assess structural stability of candidate CIHPs from point of view of ions close packing, we calculate the Goldschmidt tolerance factor *t* and the octahedral factor *μ* within the framework of idealized solid-sphere model. The statistically established empirical criteria for formability of halide perovskites is $0.81 < t < 1.11$ and $0.44 < \mu < 0.90$.[9] For the current quaternary A$_2$[BC]X$_6$ double-perovskite system the effective *t* and *μ* are defined as $t_{eff} = (R_A + R_X) / \sqrt{2}((R_B + R_C)/2 + R_X)$ and $\mu_{eff} = (R_B + R_C)/2R_X$, where *R* are Shannon ionic radii[10] and the average between $R_B$ and $R_C$ are taken as the effective radius of the octahedral-site ion. The results are summarized in Figure S12. The values of $t_{eff}$ lie in the range of 0.86~1.04, all satisfying the stable criterion of *t*. Turning to $\mu_{eff}$, we find less than half of candidate CIHPs meet the stable criterion and generally the compounds with positive $\Delta H_{dec}$ (see text) have the higher $\mu_{eff}$ (falling in or approaching the stable region).

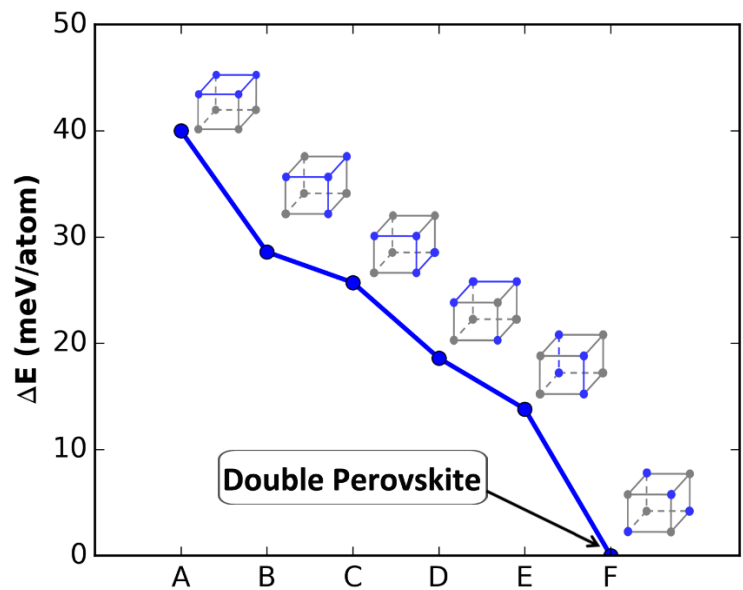

**Figure S1**. Energies of $Cs_2AgInCl_6$ composed of different types of $AgCl_6$ (in grey) + $InCl_6$ (in blue) motifs arrangements. The calculations are performed with 2x2x2 supercell of standard cubic perovskite structure. The total number of structural configurations is 6. The lowest-energy configuration **F** (set to energy zero) corresponds to the double-perovskite or elpasolite structure (in space group of *Fm-3m*), where the $AgCl_6$ and $InCl_6$ motifs alternate along the three crystallographic axes, forming the rock-salt type ordering.

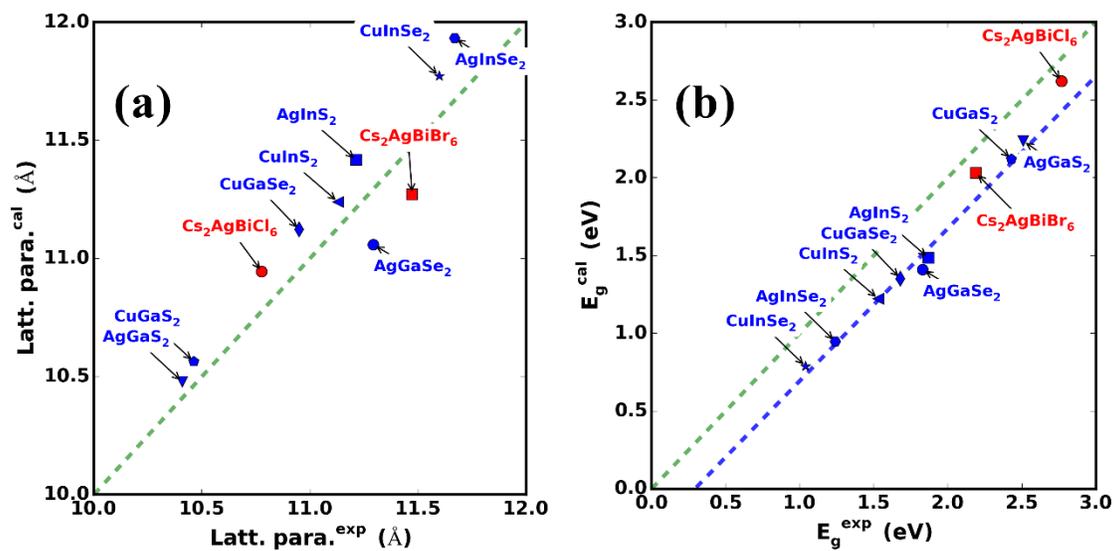

**Figure S2**. Calculated lattice parameters (a) and band gaps (b) for known chalcopyrites and Bi-based $A_2[BC]X_6$ halide perovskites, compared with experimental data. Lattice parameter *c* is taken for tetrahedral chalcopyrites. The HSE functional is used both for structural optimization and for evaluating the band gap at the optimized geometry. The experimental data are taken from Ref. 11 for chalcopyrites and Ref. 12 for Bi-based halide perovskites.

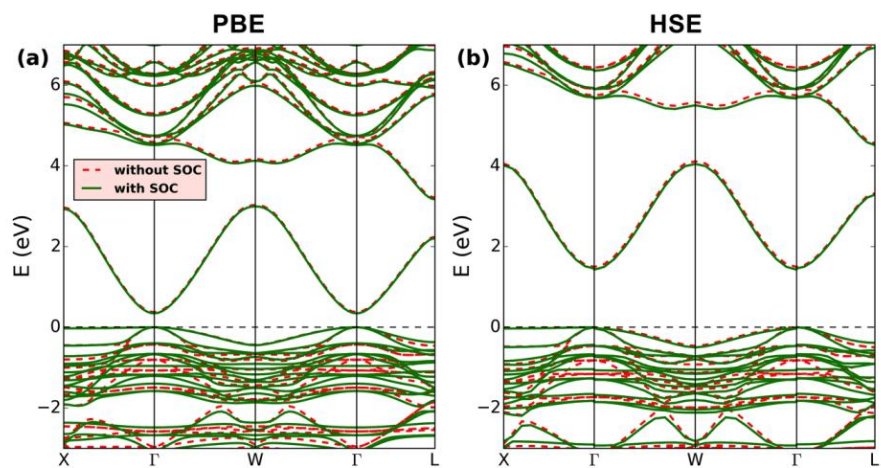

**Figure S3**. Effect of spin-orbit coupling (SOC) on band structure of $Cs_2AgInBr_6$. The results with/without inclusion of the SOC are shown by solid green/dash red lines. For both cases the valence band maximum is set to energy zero. Examination is carried out by using both the (a) PBE and (b) HSE functionals. One sees that the SOC has mild effect on the band structure of the CIHP compounds.

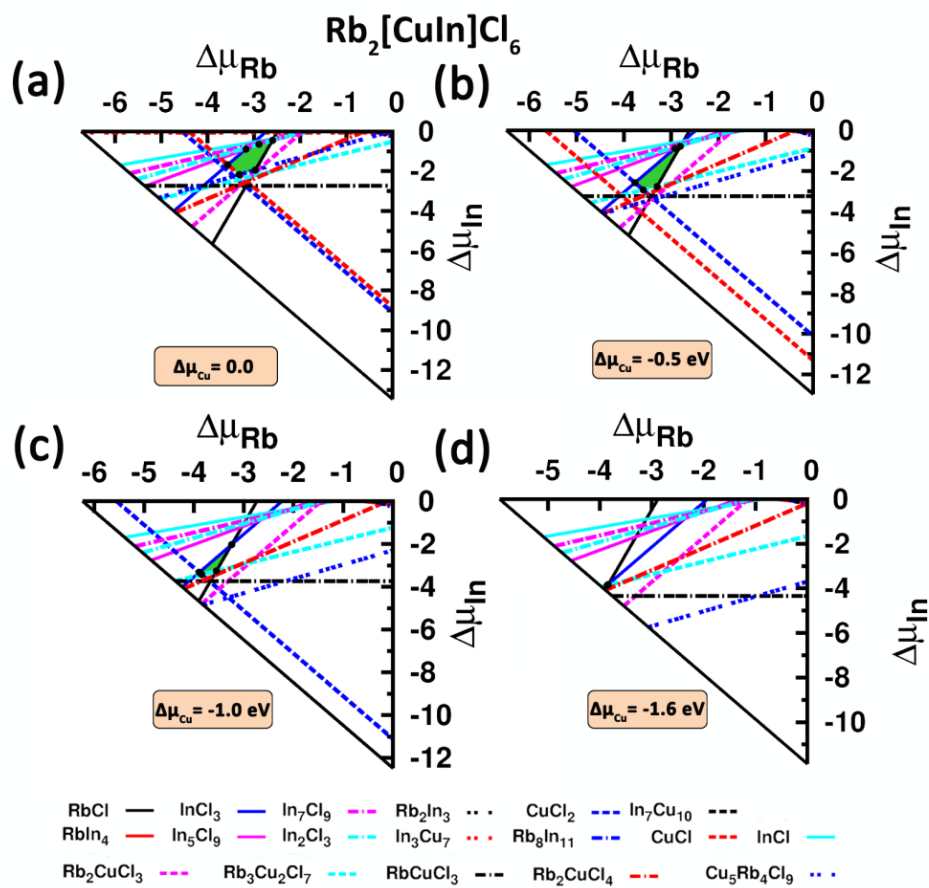

**Figure S4.** The phase stability diagram analysis (as described in Supplementary Sec. I) results sliced at several Cu-varied growth conditions represented by $\Delta\mu_{Cu}$ (see text) for $Rb_2[CuIn]Cl_6$. The polygon region in green represents thermodynamic stable condition and each line corresponds to one competing phase. The main directly competing phases, which critically control the CIHP compound stability and correspond to the lines surrounding the green stable polygon region, are $InCl_3$, $RbCl$, $CuCl_2$, $In_2Cl_3$, $CuCl$, $Rb_3Cu_2Cl_7$, and $Rb_4Cu_5Cl_9$. The last subplot (d) represents the critical condition of the thermodynamic stability, at which the green stable polygon region shrinks to a point and is disappearing.

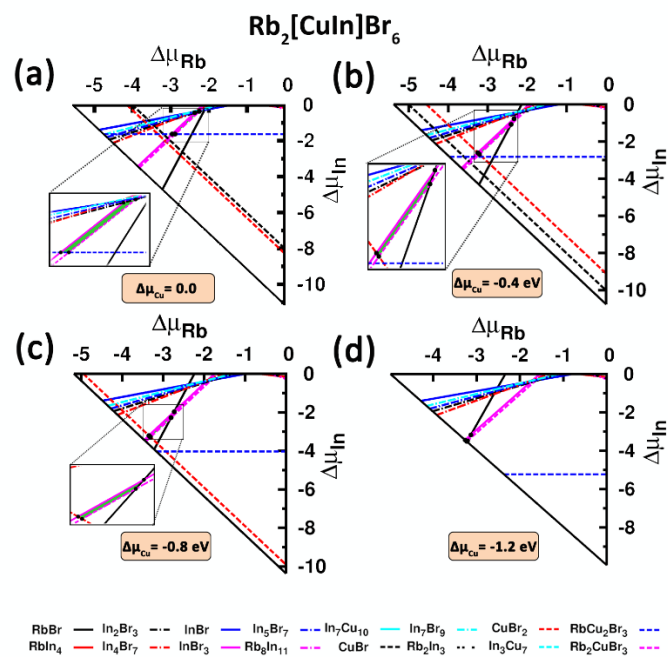

**Figure S5**. The phase stability diagram analysis results sliced at several Cu-varied growth conditions represented by $\Delta\mu_{Cu}$ (see text) for Rb$_2$[CuIn]Br$_6$. The polygon region in green represents thermodynamic stable condition and each line corresponds to one competing phase. The main directly competing phases, which critically control the CIHP compound stability and correspond to the lines surrounding the green stable polygon region, are RbIn$_4$, InBr$_3$, Rb$_2$CuBr$_3$, and RbBr. The last subplot (d) represents the critical condition of the thermodynamic stability, at which the green stable polygon region shrinks to a point and is disappearing.

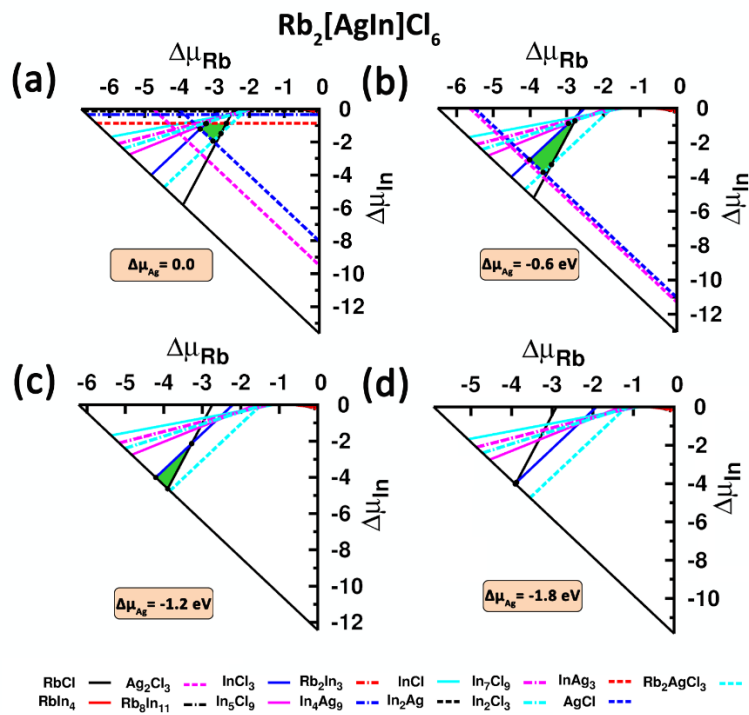

**Figure S6.** The phase stability diagram analysis results sliced at several Ag-varied growth conditions represented by $\Delta\mu_{Ag}$ (see text) for Rb$_2$[AgIn]Cl$_6$. The polygon region in green represents thermodynamic stable condition and each line corresponds to one competing phase. The main directly competing phases, which critically control the CIHP compound stability and correspond to the lines surrounding the green stable polygon region, are InCl$_3$, RbCl, AgCl, InAg$_3$, and In$_2$Cl$_3$. The last subplot (d) represents the critical condition of the thermodynamic stability, at which the green stable polygon region shrinks to a point and is disappearing.

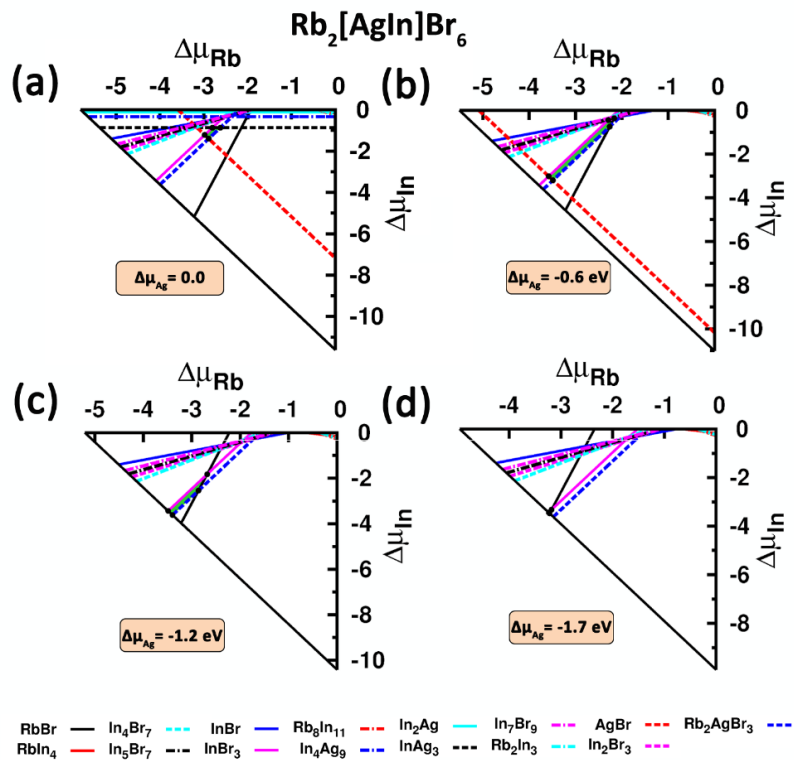

**Figure S7**. The phase stability diagram analysis results sliced at several Ag-varied growth conditions represented by $\Delta\mu_{Ag}$ (see text) for Rb$_2$[AgIn]Br$_6$. The polygon region in green represents thermodynamic stable condition and each line corresponds to one competing phase. The main directly competing phases, which critically control the CIHP compound stability and correspond to the lines surrounding the green stable polygon region, are InBr$_3$, RbBr, Rb$_2$AgBr$_3$, AgBr, and InAg$_3$. The last subplot (d) represents the critical condition of the thermodynamic stability, at which the green stable polygon region shrinks to a point and is disappearing.

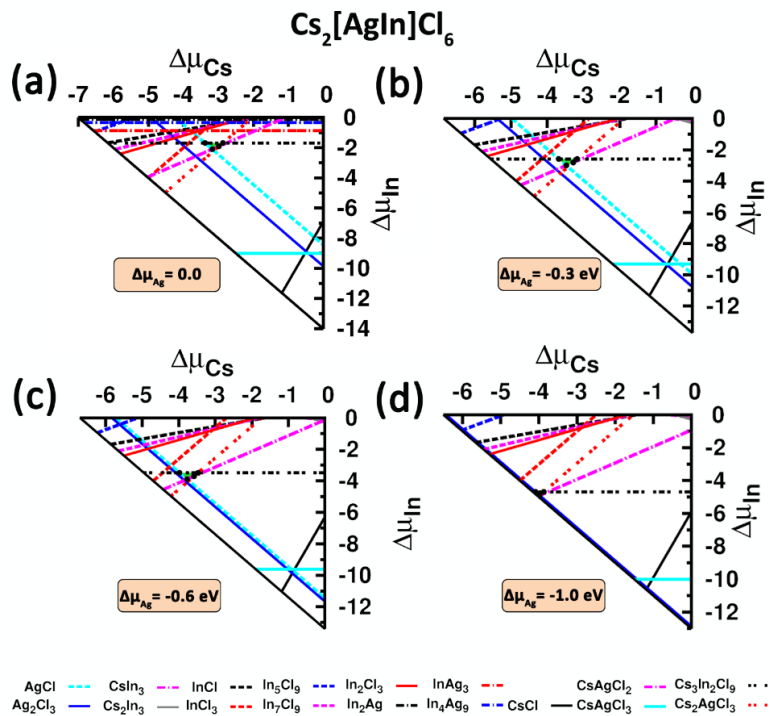

**Figure S8**. The phase stability diagram analysis results sliced at several Ag-varied growth conditions represented by $\Delta\mu_{Ag}$ (see text) for $Cs_2[AgIn]Cl_6$. The polygon region in green represents thermodynamic stable condition and each line corresponds to one competing phase. The main directly competing phases, which critically control the CIHP compound stability and correspond to the lines surrounding the green stable polygon region, are $Cs_3In_2Cl_9$, $AgCl$, $CsAgCl_2$, and $Cs_2AgCl_3$. The last subplot (d) represents the critical condition of the thermodynamic stability, at which the green stable polygon region shrinks to a point and is disappearing.

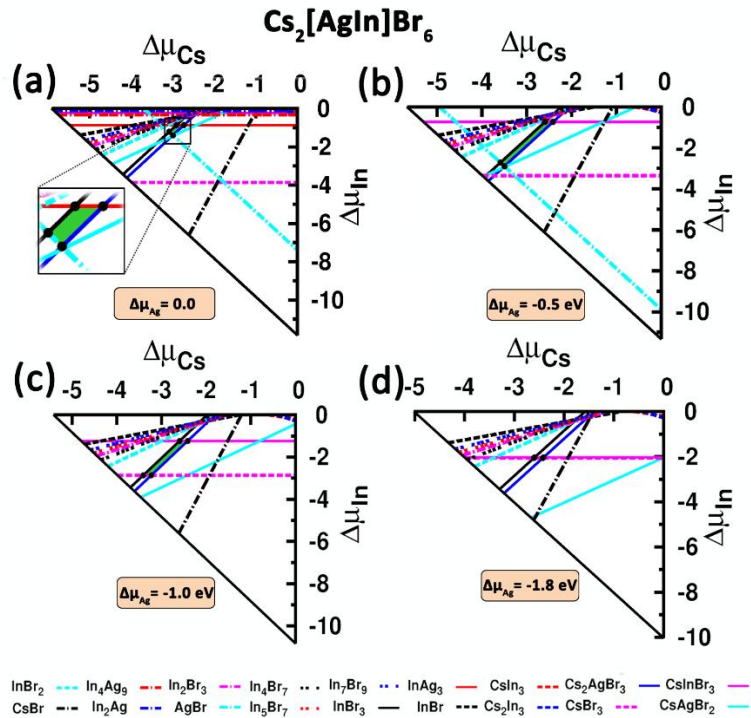

**Figure S9**. The phase stability diagram analysis results sliced at several Ag-varied growth conditions represented by $\Delta\mu_{Ag}$ (see text) for $Cs_2[AgIn]Br_6$. The polygon region in green represents thermodynamic stable condition and each line corresponds to one competing phase. The main directly competing phases, which critically control the CIHP compound stability and correspond to the lines surrounding the green stable polygon region, are $InBr_3$, $Cs_2AgBr_3$, $InAg_3$, $AgBr$, $CsAgBr_2$, and $CsInBr_3$. The last subplot (d) represents the critical condition of the thermodynamic stability, at which the green stable polygon region shrinks to a point and is disappearing.

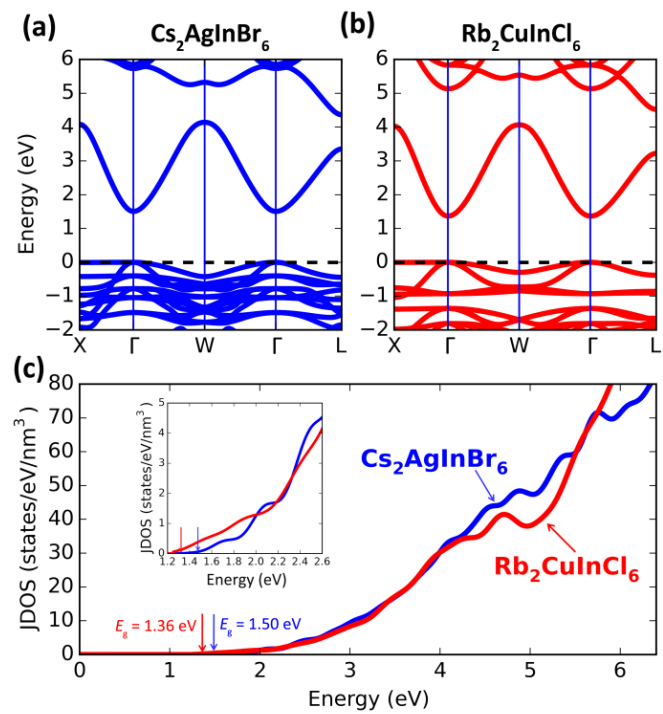

**Figure S10**. Calculated electronic band structures (a, b) and (c) joint density of states (JDOS) of $Cs_2AgInBr_6$ and $Rb_2CuInCl_6$. In (a, b) the valence band maximum is set to energy zero. The inset of (c) shows the zoomed-in plot of JDOS in proximity to the band gap.

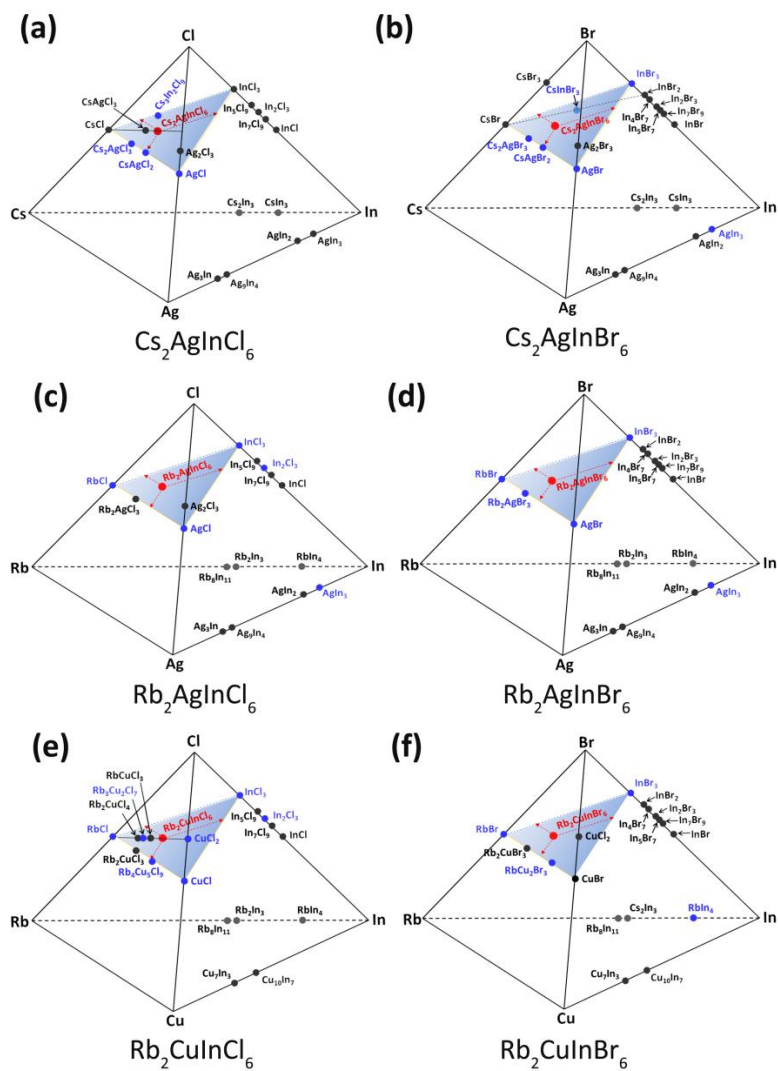

**Figure S11**. Tetrahedral phase diagram of quaternary A-B-C-X system for the six proposed optimal CIHP $A_2BCX_6$ compounds. All the known existing (binary and ternary) phases (considered by the phase stability diagram analysis in Fig. 3 of the main text and Supplementary Figs. S4-S9) are mapped into the tetrahedron with elemental compositions of A, B, C, and X as vertexes. The directly competing phases, which critically control the CIHP compound stability and correspond to the lines surrounding the green stable polygon region in Fig. 3, are shown in blue.

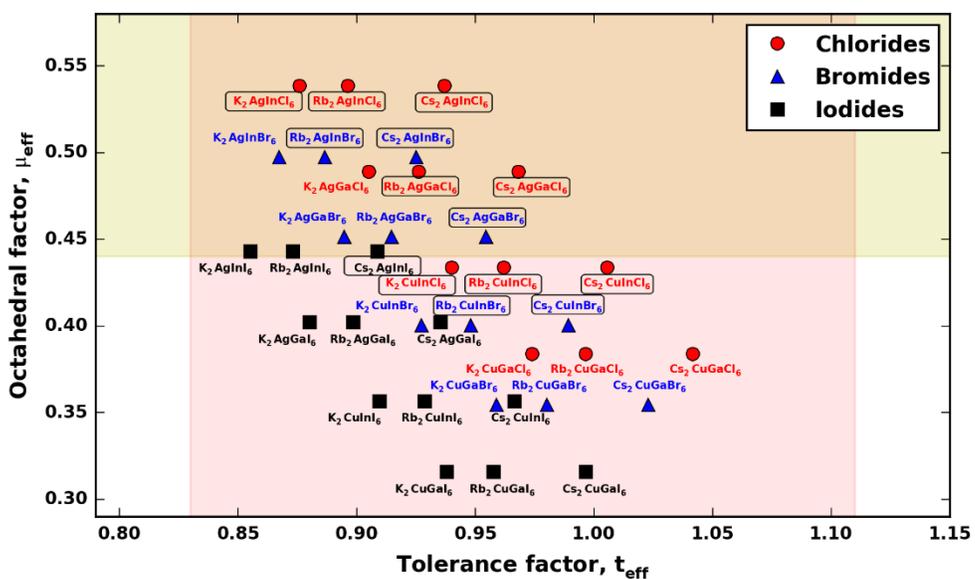

**Figure S12**. Mapping of all the candidate CIHPs onto two-dimensional plot with the effective Goldschmidt tolerance factor $t_{eff}$ and the effective octahedral factor $\mu_{eff}$ as variables. The statistically established empirical criteria for formability of halide perovskites,[9] *i.e.*, 0.81 < $t$ < 1.11 and 0.44 < $\mu$ < 0.90, is shaded. Formulas of the CIHPs with positive decomposition enthalpy ($\Delta H_{dec}$, see Table S1) are marked in rounded squares.

Table S1. Calculated explicit data of lattice parameter ($a$), band gap ($E_g$), decomposition enthalpy with respect to the disproportionation channel into binary competing phases ($\Delta H_{dec}$), thermodynamically stable condition (with respect to the disproportionation channels into all possible competing phases, see Supplementary Sec. I), carrier effective masses ($m_e^*$ for electron, $m_{hh}^*$ for heavy hole, and $m_{lh}^*$ for light hole), and exciton binding energy ($E_b$, evaluated by using the hydrogen-like Wannier-Mott exciton model) for 36 candidate CIHPs considered for materials screening. The $E_g$ is calculated by using the HSE functional with standard 25% exact Fock exchange. The effective masses and $E_b$ are calculated only for six thermodynamically stable CIHPs. For the $E_b$, two values are calculated by using $m_{lh}^*$ (the former) and $m_{hh}^*$ (the latter), respectively.

| $A_2BCX_6$ | | | | $a$ (Å) | $E_g$ (eV) | $\Delta H_{dec}$ (meV/atom) | Stable? (×/√) | $m^*$ | | | $E_b$ (meV) |
| --- | --- | --- | --- | --- | --- | --- | --- | --- | --- | --- | --- |
| A | B | C | X | | | | | $m_e^*$ ($m_0$) | $m_{lh}^*$ ($m_0$) | $m_{hh}^*$ ($m_0$) | |
| Cs | Cu | Ga | Cl | 10.107 | 1.43 | -7 | × | -- | -- | -- | -- |
| | Cu | Ga | Br | 10.660 | 0.50 | -23 | × | -- | -- | -- | -- |
| | Cu | Ga | I | -- | -- | -74 | -- | -- | -- | -- | -- |
| | Cu | In | Cl | 10.333 | 1.40 | 77 | × | -- | -- | -- | -- |
| | Cu | In | Br | 10.883 | 0.67 | 23 | × | -- | -- | -- | -- |
| | Cu | In | I | -- | -- | -22 | -- | -- | -- | -- | -- |
| | Ag | Ga | Cl | 10.369 | 2.56 | 40 | × | -- | -- | -- | -- |
| | Ag | Ga | Br | 10.914 | 1.32 | 13 | × | -- | -- | -- | -- |
| | Ag | Ga | I | -- | -- | -51 | -- | -- | -- | -- | -- |
| | Ag | In | Cl | 10.594 | 2.52 | 116 | √ | 0.32 | 0.43 | 2.38 | 195/304 |
| | Ag | In | Br | 11.156 | 1.50 | 56 | √ | 0.24 | 0.34 | 1.37 | 97/139 |
| | Ag | In | I | 11.962 | 0.31 | 3 | -- | -- | -- | -- | -- |
| K | Cu | Ga | Cl | 9.848 | 1.30 | -25 | × | -- | -- | -- | -- |
| | Cu | Ga | Br | 10.462 | 0.41 | -72 | × | -- | -- | -- | -- |
| | Cu | Ga | I | -- | -- | -133 | -- | -- | -- | -- | -- |
| | Cu | In | Cl | 10.135 | 1.35 | 36 | × | -- | -- | -- | -- |
| | Cu | In | Br | 10.724 | 0.62 | -42 | × | -- | -- | -- | -- |
| | Cu | In | I | -- | -- | -91 | -- | -- | -- | -- | -- |
| | Ag | Ga | Cl | 10.164 | 2.44 | -7 | × | -- | -- | -- | -- |
| | Ag | Ga | Br | 10.745 | 1.22 | -55 | × | -- | -- | -- | -- |
| | Ag | Ga | I | -- | -- | -120 | -- | -- | -- | -- | -- |
| | Ag | In | Cl | 10.464 | 2.48 | 51 | × | -- | -- | -- | -- |
| | Ag | In | Br | 11.023 | 1.44 | -25 | × | -- | -- | -- | -- |
| | Ag | In | I | 11.806 | 0.26 | -75 | -- | -- | -- | -- | -- |
| Rb | Cu | Ga | Cl | 9.940 | 1.34 | -5 | × | -- | -- | -- | -- |
| | Cu | Ga | Br | 10.532 | 0.43 | -25 | × | -- | -- | -- | -- |
| | Cu | Ga | I | -- | -- | -107 | -- | -- | -- | -- | -- |
| | Cu | In | Cl | 10.237 | 1.36 | 65 | √ | 0.30 | 0.63 | 3.40 | 123/191 |
| | Cu | In | Br | 10.808 | 0.63 | 11 | √ | 0.18 | 0.32 | 1.75 | 81/122 |
| | Cu | In | I | -- | -- | -60 | -- | -- | -- | -- | -- |
| | Ag | Ga | Cl | 10.204 | 2.48 | 24 | × | -- | -- | -- | -- |
| | Ag | Ga | Br | 10.781 | 1.26 | -1 | × | -- | -- | -- | -- |
| | Ag | Ga | I | -- | -- | -89 | -- | -- | -- | -- | -- |
| | Ag | In | Cl | 10.520 | 2.50 | 89 | √ | 0.32 | 0.41 | 2.35 | 234/327 |
| | Ag | In | Br | 11.064 | 1.46 | 34 | √ | 0.24 | 0.32 | 1.81 | 75/100 |
| | Ag | In | I | 11.901 | 0.27 | -40 | -- | -- | -- | -- | -- |

**References:**

(1) Heyd, J.; Scuseria, G. E.; Ernzerhof, M. *J. Chem. Phys.* **2003**, *118*, 8207–8215.

(2) Persson, C.; Zhao, Y.-J.; Lany, S.; Zunger, A. *Phys. Rev. B* **2005**, *72*, 35211.

(3) Li, Y.; Singh, D. J.; Du, M.-H.; Xu, Q.; Zhang, L.; Zheng, W.; Ma, Y. *J. Mater. Chem. C* **2016**, *4*, 4592–4599.

(4) Togo, A.; Tanaka, I. *Scr. Mater.* **2015**, *108*, 1–5.

(5) Souvatzis, P.; Eriksson, O.; Katsnelson, M. I.; Rudin, S. P. *Phys. Rev. Lett.* **2008**, *100*, 95901.

(6) Gajdoš, M.; Hummer, K.; Kresse, G.; Furthmüller, J.; Bechstedt, F. *Phys. Rev. B* **2006**, *73*, 45112.

(7) Yu, L.; Zunger, A. *Phys. Rev. Lett.* **2012**, *108*, 68701.

(8) Yu, L.; Kokenyesi, R. S.; Keszler, D. A.; Zunger, A. *Adv. Energy Mater.* **2013**, *3*, 43–48.

(9) Li, C.; Lu, X.; Ding, W.; Feng, L.; Gao, Y.; Guo, Z. *Acta Crystallogr. B* **2008**, *64*, 702–707.

(10) Shannon, R. D. *Acta Crystallogr. Sect. A* **1976**, *32*, 751–767.

(11) Jaffe, J. E.; Zunger, A. *Phys. Rev. B* **1984**, *29*, 1882–1906.

(12) McClure, E. T.; Ball, M. R.; Windl, W.; Woodward, P. M. *Chem. Mater.* **2016**, *28*, 1348–1354.